\documentclass[printer]{aa} 

\usepackage{graphicx}
\usepackage{txfonts}
\usepackage{natbib}
\usepackage{color}
\usepackage{lscape}

\def\fdeg{\hbox{$^\circ$}}

\usepackage{amstext}

\begin{document}
\title{Towards a 3D kinetic tomography of Taurus clouds: \thanks{Based on observations obtained at the T\'elescope Bernard Lyot (TBL) at Observatoire du Pic du Midi, CNRS/INSU and Universit\'e de Toulouse, France.}}

\subtitle{I - Linking neutral potassium and dust}

\author{A. Ivanova\inst{1,2}
\and
R. Lallement\inst{3}
\and
J.L. Vergely\inst{4}
\and
C. Hottier\inst{3}
}

\institute{LATMOS, Universit\'e Versailles-Saint-Quentin, 11 Bd D'Alembert, Guyancourt, France\\
              \email{anastasia.ivanova@latmos.ipsl.fr}
  \and
 Space Research Institute (IKI), Russian Academy of Science, Moscow 117997, Russia 
  \and
  GEPI, Observatoire de Paris, PSL University, CNRS,  5 Place Jules Janssen, 92190 Meudon, France
  \and
  ACRI-ST, 260 route du Pin Montard, 06904, Sophia Antipolis, France
}

\date{Received ; accepted }
\titlerunning{Taurus 3D kinetic maps}
 
\abstract
{Gaia parallaxes and photometric measurements open a 3D era for the Milky Way, including its interstellar (IS) matter. 3D Galactic dust distributions are constructed in various ways, based on Gaia data and photometric or spectroscopic surveys.}
{The assignment of radial motions to IS dust structures seen in 3D, or 3D kinetic tomography, would be a valuable tool allowing to connecting the structures to emission lines of the associated gas, now measured at increasingly higher spectral and angular resolutions and rich of information on physical and chemical processes. To this end, one of potential technique is to establish a link between dust clouds and Doppler velocities of absorption lines imprinted in stellar spectra by the gas associated with the dust. This requires a relatively close correlation between the absorber column and the dust opacity. We have investigated the link between the strength of interstellar K\,{\sc i} absorption and the opacity of the dust in front of stars in the Taurus area, and tested the feasibility of assigning velocities to 3D dust clouds, on the basis of K\,{\sc i}  absorption data.}
{We have obtained high spectral resolution and high signal to noise spectra of 58 early-type stars in the direction of the Taurus, Perseus and California molecular clouds. We have developed a new, dual interstellar and telluric profile-fitting technique to extract the interstellar K\,{\sc i} 7665, 7699 A absorption lines from stellar spectra and applied it to the new data and to archived spectra of 58 additional targets. In parallel, we have updated 3D dust maps reconstructed through inversion of individual stellar light extinctions. To do so, we supplemented the catalog of extinction estimates based on Gaia and 2MASS photometry with recently published extinction catalogs based on stellar spectroscopic surveys.   We used  the 3D map and the set of velocity components seen in absorption to assign radial velocities to the dust clouds distributed along their paths in the most consistent way.}
{We illustrate our profile-fitting technique and present the K\,{\sc i} velocity structure of the dense ISM along the paths to all targets. As a validation test of the dust map, we show comparisons between distances to several reconstructed clouds with recent distance assignments based on different techniques. Target star extinctions estimated by integration in the 3D map are compared with their K\,{\sc i} 7699 A absorptions and the degree of correlation is found comparable to the one between the same K\,{\sc i} line and the total hydrogen column for stars distributed over the sky that are part of a published high resolution survey.  We show images of the updated dust distribution in a series of vertical planes in the Galactic longitude interval 150-182.5 deg  and our estimated assignments of radial velocities to the opaque regions. Most clearly defined K\,{\sc i} absorptions may be assigned to a dense dust cloud between the Sun and the target star. It appeared relatively straightforward to find a velocity pattern consistent will all absorptions and ensuring coherence between adjacent lines of sight, at the exception of a few weak lines. We compare our results with recent determinations of velocities of several clouds and find good agreement. These results demonstrate that the extinction-K\,{\sc i} relationship is tight enough to allow linking the radial velocity of the K\,{\sc i} lines to the dust clouds seen in 3D, and that their combination may be a valuable tool in building a 3D kinetic structure of the dense ISM. We  discuss limitations and perspectives for this technique.}
{} 


\maketitle

%

\section{Introduction}
The history and present evolutionary state of the Milky Way are currently being deciphered at all spatial scales, boosted by the remarkable Gaia measurements \citep{Brown20}, and complemented by massive photometric and spectroscopic stellar surveys \citep[for a recent review of results on the Milky Way history see][]{Helmi20}. On the large scale, and in addition to astrometric measurements, Milky Way evolutionary models also benefit from new, accurate photometric data and subsequent realistic estimates of stellar parameters and extinction of starlight by dust. New, massive measurements of extinctions and faint sources photometric distances have been produced  \citep[e.g.,][]{Sanders18,Anders19,Queiroz20}. This has favored the development of catalogs of dust cloud distances and three-dimensional (3D) dust distributions, \citep{Green19,Chen19,Lallement19,Rezaei20,Zucker20,Leike20, Guo20, Hottier20}. On the smaller scale of star-forming regions, the combination of accurate distances and proper motions with realistic age determinations marks a new era for their detailed study \citep[see, e.g. recent studies of][]{Grossschedl21, Roccatagliata20, Kounkel18,Galli19}.

The assignment of a radial velocity to each structure represented in 3D would allow to go one step further in a number of on-going analyses, and, again, in a wide range of spatial scales. Sophisticated N-body/hydrodynamical simulations of the stellar-gaseous Galactic disc are being developed \citep[see, e.g.][and references therein]{Khoperskov20} and used to reproduce the 6D phase-space  (stellar positions and velocities) distributions provided by Gaia.  Exciting debates are currently held about the respective roles of external perturbations due to dwarf galaxy passages or internally driven resonances  associated with bar and spiral arms \citep{Antoja18, Khoperskov21}. In parallel, chemo-dynamical models are developed and aim at reproducing the observed enrichment sequences and dichotomies in  the distributions of elemental compositions \citep[e.g.][]{Haywood19,Khoperskov21, Sharma20,Katz21}. Up to now efforts have been concentrated on comparisons with the newly observed spatial distributions and chemical and dynamical characteristics of the stellar populations. However, stellar history and IS matter evolution are tightly coupled and additional constraints on the models may be brought by IS matter distribution and dynamics. On the smaller scale of star-forming regions, dense IS matter and stars are dynamically coupled, as was recently quantified by \cite{Galli19} who found a difference of less than 0.5 km s$^{-1}$ between the radial velocities of young stellar objects (YSOs) and the associated CO in Main Taurus regions. This shows that the velocity distribution of the interstellar gas and dust may help reconstructing the star formation history. Extremely detailed spectro-imaging surveys have been performed or are in progress whose aims are state-of-the-art studies of physical and chemical processes at work in the various phases of the ISM in such regions \citep[see, e.g.][]{Pety17}.  The assignment of  all emission lines, which contain very rich information on the processes, to their source regions, can help constraining and refining the models. Finally, and for all interstellar medium studies, the assignment of velocities to interstellar clouds located in 3D would allow connecting the structures to their multi-wavelength emission through velocity cross-matching. Extended 3D kinetic maps would display multiple structures sharing the same radial velocity and located at different distances, if any, and be useful to clarify the models.

Here we use the term 3D kinetic tomography for the assignment of velocities to structures occupying a volume in 3D space, i.e., it does not include the association of absorption and emission lines along the same individual line of sight. A structure represented in 3D may be a dust cloud reconstructed by inversion, a voxel in the case of discretized 3D maps of dust or gas, or by extension a cloud having some extent in 2D images and localized in distance.  Several recent works have been devoted to 3D kinetic tomography. \cite{Tchernyshyov17} developed a method using HI and CO spectral cubes (i.e. position-position-velocity matrices) on the one hand and a 3D reddening map (i.e. a position-position-position matrix) on the other hand. The authors adjusted the radial velocity in each voxel of a discretized volume around the Plane until achievement of consistency between the three datasets, assuming conversion factors between CO, HI temperatures and dust opacity.  \cite{Tchernyshyov18} used the series of measurements of a NIR diffuse interstellar band (DIB) in SDSS/APOGEE stellar spectra and  photometric distances of the target stars to derive a planar map of the radial velocity. The authors used variations in measured DIB absorption profiles for stars at slightly different distances and directions to extract the contribution of the local interstellar matter. \cite{Zucker18} used the distance-reddening posterior distributions from the Bayesian technique of \cite{Green18} and $^{12}$CO spectra to assign radial velocities to the main structures in Perseus. To do so, the authors associated each opacity bin in Perseus with a linear combination of velocity slices.

 3D kinetic tomography is not straightforward, and one can reasonably foresee that using different tracers and different techniques will help achieving accurate results in large volumes. The difficulties are of various types. The omni-directional spatial resolution of 3D dust maps computed for large volumes has not yet fully reached a level that allows to identify the parsec or sub-parsec counterparts of extremely small details in the ultra high angular and spectral resolution radio data. The Pan-STARRS/2MASS map of \cite{Green19} has a very high angular resolution, similar to the one of emission data,  however the resolution along the radial direction is much poorer than in the plane of the sky. The hierarchical technique presented in \cite{Lallement19} was still limited to a 25 pc wide 3D kernel, and the updated map presented in the present article and based on the same technique is limited to 10 pc resolution. The method used by \cite{Leike20} allowed the authors to reach 1 pc spatial resolution, however for a limited volume due to the computational cost. It is hoped that this difficulty should gradually decrease thanks to future releases of Gaia catalogs and the extension of ground surveys. If emission spectra are the sources of the Doppler velocities, a second difficulty is the existence of clouds at different distances and sharing the same radial velocity, a degeneracy that increases sharply at low latitudes. The techniques of \cite{Tchernyshyov17}  and \cite{Zucker18}  may, in principle, disentangle the contributions, however this requires prior knowledge of the ratio between the dust opacity and the species used in emission. E.g., \cite{Zucker18} assumed that the dust located closer than 200 pc from the Sun in front of IC348 in Perseus does not contribute to the CO emission, despite the non-negligible ($\geq$ 35\%) contribution to the extinction of a foreground cloud at 170 pc (see Fig. \ref{fig:lon3} for the visualization of this foreground cloud). One way to help breaking the degeneracy is the additional use of distance-limited IS absorption data. In this case, target stars must be distributed within and beyond the volume containing the main IS cloud complexes. Finally, if extinction maps are part of the tomography, an additional difficulty is due to the fact that dust traces both dense molecular and atomic gas phases, and, as a consequence, emission or absorption data used for the Doppler shift measurements must trace these two phases. This is illustrated in Figure 1 that displays $^{12}$CO columns in the Taurus area and superimposed HI 21 cm intensity contours, both for the LSR velocity range of the local IS matter (here $-10 \leq V \leq +10$ km s${-1}$). The marked and well-known differences between the respective amounts of molecular and atomic gas clearly show that using solely one of these two species for a cross-identification with dust structures everywhere in the image, is inadequate, unless one restricts the study to predominantly molecular regions, as in \cite{Zucker18}, in which case CO is a convenient unique tracer.

 The objective of the present work is to test the use of neutral potassium (K\,{\sc i}) absorption in 3D kinetic ISM tomography. Absorption by interstellar neutral potassium is expected to allow both dense atomic and molecular phases to be traced, according to the results of the extensive study devoted to this species by \cite{WeltyHobbs01} (herafter WH01). Using a large number of Galactic K\,{\sc i} measurements, extracted from very high resolution spectra, the authors studied the link between  K\,{\sc i}  and various species, and found a quadratic dependence of  K\,{\sc i}  with hydrogen (H$_{tot}$= HI +2 H$_{2}$). Because H$_{tot}$ is known to correlate with the extinction, we may expect the a similar relationship between neutral potassium and extinction.  Note that according to the WH01 study,  the K\,{\sc i} - H$_{tot}$ correlation is better than for Na\,{\sc i} with H$_{tot}$, which favors the use of the former species. Additional advantages of using potassium by comparison with neutral sodium is its higher mass, hence its narrower and deeper absorption lines, more appropriate to studies of the velocity structure. In comparison with DIBs, the very small width of the  K\,{\sc i} lines allows a much better disentangling of multiple velocity components and significantly higher accuracy of the absolute values for each one. The caveat is that, K\,{\sc i} being detected in the optical, highly reddened stars are not part of potential targets and one must restrict observations to the periphery of the dense clouds. Using NIR DIBs, on the contrary, allows to use targets located behind high opacity clouds.

 Our test consisted in -gathering high spectral resolution, high quality stellar spectra of target stars located in front of, within and behind the Taurus, Perseus and California clouds,- extracting from the spectra interstellar K\,{\sc i} absorption profiles and measuring the radial velocities of the absorptions lines, -updating 3D extinction maps to achieve better spatial resolution, -comparing the K\,{\sc i} absorption strengths with the integrated extinction along the path to the targets, -attempting a synthetic assignment of the measured  K\,{\sc i} radial velocities to the clouds contributing to this extinction, -evaluating the extent of information on the velocity pattern that can be deduced from K\,{\sc i} data alone.

Neutral potassium is detectable by its  $\lambda\lambda$7665, 7699 \AA~ resonance doublet. Unfortunately, the first, stronger transition is located in the central part of the A-band of telluric molecular oxygen. This is why the second, weaker transition is used alone or the data must be corrected for the telluric absorption.  Here we present a novel method to extract a maximum of information from the two transitions that avoids using telluric-corrected spectra obtained by division by a spectrum of a standard star or by a model. Such corrected spectra are often characterized by strong residuals in the regions of the deepest telluric lines that make difficult the determination of the continuum and the subsequent profile-fitting. To avoid this, we use two distinct, consecutive adjustments. First, we use the atmospheric transmittance models from the  TAPAS on-line facility \footnote{http://cds-espri.ipsl.fr/tapas/} \citep{Bertaux14}, and select those that are suitable for the respective observing sites. We use these models and their adjustments to the data to (i) refine the wavelength calibration, (ii) determine the instrumental function (or line spread function, hereafter LSF) as a function of wavelength along the echelle order and (iii) derive the atmospheric transmittance that prevailed at each observation, prior to entry into the instrument. In a second step, this adjusted transmittance and the spectral resolution measurements are used in a global forward model in combination with a multi-cloud model of the two IS K\,{\sc i} 7665 and 7699 \AA{} transitions. 

The spectral data were recorded with the Pic du Midi TBL-Narval spectrograph during two dedicated programs  and are complemented by archival data from the Polarbase facility, recorded with Narval or the ESPaDOnS spectro-polarimeter at CFHT \citep{Donati97,Petit14} and several other archival data.  The extracted K\,{\sc i} lines and their radial velocities are compared with the locations of the dust clouds in an updated 3D dust map. This new 3D distribution has been derived from the inversion of a catalog of individual extinction measurements that comes out as an auxiliary result of the \cite{Sanders18} extensive analysis devoted to stellar populations ages, a study based on 6 massive spectroscopic surveys. This new Bayesian inversion used as a prior the 3D distribution recently derived from Gaia and 2MASS \citep{Lallement19}. 

The article is organized in the following way. Spectral data are presented in section 2.  The telluric modeling is then described in section 3, and the K\,{\sc i} extraction method is detailed in section 4. In section 5 we describe the construction of the updated 3D dust map and test the link between the integrated extinction along the path to the star and the K\,{\sc i}  absorption intensity. Section 6 shows the velocity assignments to the dense structures that were found compatible with the whole dataset and appeared as the most coherent ones. In the last section we discuss the results and the perspectives of such an approach, as well as its future association with the information from emission data.

\section{Spectral data}

\subsection{TBL Narval data}
High resolution, high signal spectra were recorded in service mode with the Narval Echelle spectro-polarimeter at the Bernard Lyot Telescope (TBL) facility during two dedicated programs (L152N07 and L162N04, P.I. Lallement). The Narval spectra cover a wide wavelength range from 370 to 1058 nm. Target stars were selected in direction and distance to sample the volumes of Taurus, Perseus and California clouds, and were chosen preferentially among the earliest types. The observations were done in the "star only" (pure spectroscopic) mode that provides a resolving power on the order of 75,000. Most of the observations were performed at low airmass.  

Spectra of good and excellent quality  were obtained for a total of 57 new targets listed in Table \ref{tab:targettable}. We used the fully reduced and wavelength-calibrated spectra from the Narval pipeline. The pipeline provides the spectrum in separate echelle orders, and we used here the order that covers the $\simeq$ 764.5-795.0 nm interval and contains the two K\,{\sc i} transitions at 7664.911 and 7698.974 \AA. The first transition is also included in an adjacent order, however we used a unique order and non re-sampled data for the K\,{\sc i} extraction. We discuss our reasoning in section 3. This is also valid for the archive Narval and ESPaDOnS data described below. The whole spectrum was used to check for the spectral type, the stellar line widths and the interstellar Na\,{\sc i} absorption lines (see section 4). Further studies of the full set of atomic and molecular absorption lines and of the diffuse interstellar bands are in progress. 

\subsection{Archival data}

\begin{figure}
\centering
\includegraphics[width=0.95\hsize]{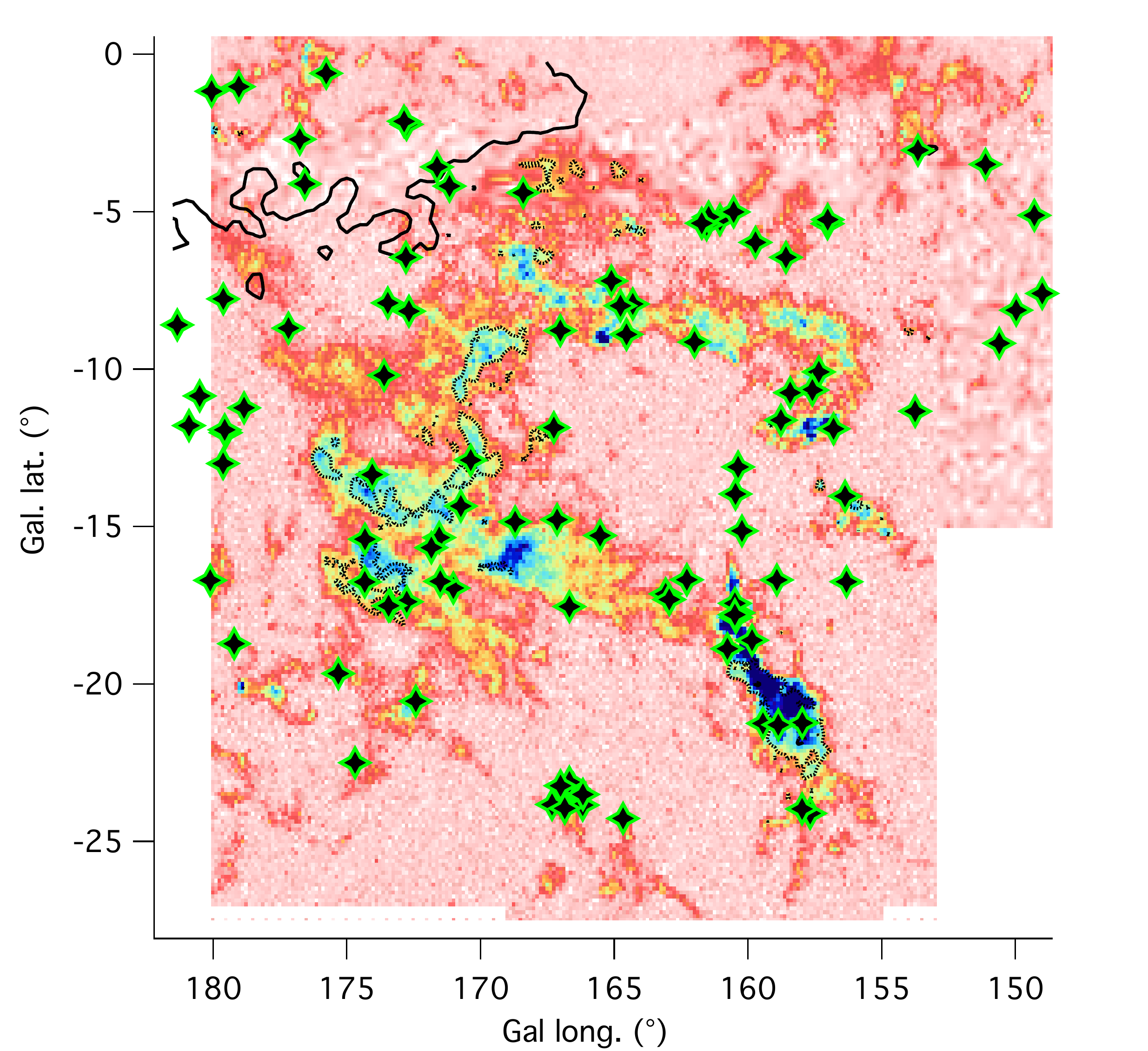}
\caption{Target stars from our program and from archives superimposed on a $^{12}$CO map of the Taurus-Perseus-California area \citep{Dame2001}. Superimposed are HI iso-contours for T(HI)= 1 , 2 and 3.5 K (thin, dot-dashed and thick line lines), based on \cite{HI4PI16} data. Both CO and HI columns were restricted to LSR velocities between -10 and +10 km s$^{-1}$.}
\label{targets}
\end{figure}

We performed an online search in the POLARBASE database and found 10 Narval  and 15 ESPaDOnS spectra of target stars that have been observed as part of other programs and are suitable for our study. Their spectral resolution R is comprised between $\simeq55000$ and $\simeq85000$, with the the lower value for the polarimetric mode. 
We also searched for useful spectra in the ESO archive facility and found spectra for 7 UVES, 2 XSHOOTER and 1 FEROS targets of interest (R= $\simeq$70000, 18000 and 80000 respectively). We also used 6 spectra from the Chicago database recorded with the ARCES spectrograph at R=$\simeq31000$ \citep{Wang03_ARCES,Fan19}. Finally, one spectrum recorded with the OHP 1.93m Elodie spectrograph R$\simeq$48000 has been added. Elodie does not cover the K\,{\sc i} doublet, however we derived the Doppler velocity of the main absorbing cloud from neutral sodium lines (5890 \AA~ doublet). All targets and corresponding sources are  listed in Table \ref{tab:targettable}. Heliocentric wavelength scales were either directly provided by the archive facility or derived from the observing date in case of spectra released in the topocentric reference frame. In the special case of ARCES data we used the observing date and the telluric features. Finally, in addition to these archival data that entered our telluric correction and profile fitting, we complemented the observation list by \cite{Chaffee82} published results on Doppler velocities of Taurus clouds for 8 targets, also based on neutral potassium lines and high resolution observations. One additional result by \cite{Welty94} based on sodium lines has also been included. We did not include the WH01 result on HD 23630 ($\eta$Tau) and its measured K\,{\sc i} radial velocity +7.9 km s$^{-1}$  due to the target imprecise distance and to redundancy with HD23512 for which we infer a similar velocity of +7.3 km s$^{-1}$. The locations of the set of targets are shown in Fig. \ref{targets}, superimposed on a $^{12}$CO map \citep{Dame2001}, here restricted to LSR velocities between -10 and +10 km s$^{-1}$. A few targets are external to the map and are not shown. Also added are HI 21 cm iso-contours for the same velocity range \citep{HI4PI16}. 

\section{Spectral analysis}

\subsection{The dual  interstellar and telluric profile-fitting}
It is well known that fitting simultaneously two different transitions of the same absorbing species with different oscillator strengths gives additional constraints on the characteristics of the intervening clouds, and helps disentangling their respective contributions in the frequent case of overlapping lines. In the case of K\,{\sc i}, the stronger transition is located in a spectral region strongly contaminated by telluric O$_{2}$, and taking benefit of this transition requires to take the O$_{2}$ absorption profile into account. While, obviously, nothing can be said about the interstellar contribution to the absorption at the bottom of a fully saturated O$_{2}$ line, in most cases the interstellar K\,{\sc i} absorption or a fraction of it falls in unsaturated parts and contains information. Here we have devised a novel method that allows to extract all available information contained in the two transitions without using the division by a standard star spectrum or a telluric model prior to the extraction of the interstellar lines. This division lefts strong residuals that makes the profile-fitting particularly difficult in the spectral region of the stronger transition and we want to avoid it. The method is made of two steps. The first step consists in fitting a telluric model to the data and determining the wavelength-dependent LSF. During this step the spectral regions around the K\,{\sc i} lines are excluded. The second step is the dual profile-fitting itself. The previously derived telluric model is multiplied by as many Voigt profiles as necessary to represent the interstellar K\,{\sc i} absorption lines and the velocity distribution of intervening clouds, and the total product is convolved by the wavelength-dependent LSF. The number of clouds and the interstellar parameters are optimized to fit this convolved product to the data.

\subsection{The preliminary telluric correction and determination of the wavelength-dependent instrumental width}
For the telluric correction we used the TAPAS online facility: TAPAS uses vertical atmospheric profiles of pressure,
temperature, humidity, and most atmospheric species, interpolated in the meteorological field of the European Centre for
Medium-range Weather Forecasts (ECMWF), the HIgh-Resolution TRANsmission molecular absorption database (HITRAN) and the
Line-by-Line Radiative Transfer Model (LBLRTM) software to compute the atmospheric transmission at ultra-high resolution
(wavelength step $\simeq$ 4 10$^{-6}$ \AA ). See \cite{Bertaux14} for details about TAPAS products and all references. Here we
used TAPAS calculations for the observatory from which each spectrum under analysis was obtained. We extracted from the star
spectrum and the TAPAS model a wavelength interval of  80 \AA~ containing the K\,{\sc i} doublet, for which we performed the
following steps.
At first we created a Gaussian LSF for an estimated preliminary resolution, adapted to the central wavelength of the correction
interval. We then convolved by this LSF the initial TAPAS transmission model for one airmass unit, and through cross-correlation
we obtained a preliminary Doppler shift between the observed spectrum (data) and this model (convolved TAPAS). To do so, we
exclude the potassium lines regions from the Doppler shift estimate, using a mask with null value from 7663 to 7667 and from
7697 to 7701 for both data and model. We then used a comprehensive list of un-blended or weakly blended telluric $O_2$  lines
located in the $\simeq$ 80 \AA\ interval containing the K\,{\sc i} doublet  to obtain a first estimate of the airmass at
observation as well as a more refined   estimate of the spectral resolution. To do so, Gaussian profiles are fitted in both the
stellar and earth transmission spectra at the locations of all these potentially useful lines and the results are compared
one-by-one. Undetected too weak lines and/or lines with too noisy profiles in the data are removed from the list. The strongest
lines are selected and all corresponding ratios between data and model equivalent widths are computed. The average ratio provides a preliminary airmass (or relative optical thickness w.r.t. the initial model), and the line-width increase between
model and data provides an estimate of the mean resolution.\\
After all these preparatory steps the main correction procedure starts. It can be decomposed into 3 main steps:
\begin{enumerate}
\item First fitting to the data of the telluric model by means of the {\it rope-length} method (see below), excluding from
the spectrum the central areas of the strong telluric lines (where the flux is close or equal to zero). This step assumes a
unique Doppler and a unique Gaussian LSF. The free parameters are the airmass factor, the Doppler shift and the LSF width.
\item Computation of a decontaminated spectrum by division of the data by the adjusted transmission model, and replacement
of the ratio by an interpolated polynomial at the locations of the strong telluric lines. This provides a {\it quasi-continuum} without strong residuals. By {\it quasi-continuum}, we mean here the telluric-free spectrum, i.e. the
stellar spectrum and its interstellar features.
\item Series of fits of the convolved product of the {\it quasi-continuum} and a TAPAS transmission to the data, gradually removing constraints on the width of the instrumental profile and on the Doppler shift, and allowing their variability
within the fitted spectral interval, along with the airmass factor variability.
\end{enumerate}
In a more detailed way:\\
For the first and second steps, we start with the preliminary resolution estimated from the linewidths, and the preliminary $O_2$ column. We then run an optimization process for a freely varying $O_2$ column (or equivalently a free airmass) and a freely varying LSF width, the convergence criteria being the minimal {\it length} of the spectrum obtained after division of the data by the convolved transmission spectrum. Here the {\it length} is simply the sum of distances between consecutive data points. Such a technique uses the fact  that the minimum length corresponds to the smoothest curve and therefore to minimum residuals after the data to model ratio, i.e. a good agreement on all the line shapes. Note that this technique is sufficient for weak to moderate lines, but is applied here only as an intermediate solution \citep[see e.g. the different situations in][] {Cox17,Cami18}. 
The third step is adapted to strong lines and corresponds to a forward modeling. We assume that the data, after division by the already well determined atmospheric transmission model, provide the telluric-free continuum, at least outside the strong lines centers, and that the interpolated polynomials may represent at first order the telluric-free data everywhere. We then perform a series of fits to the data of convolved products of this adopted continuum by a transmission model,  masking the interstellar K\,{\sc i} regions. 
We let the Doppler shift free to vary along the spectral interval, then, in a final stage, we allow for a variation of the LSF width. We found that this last stage is very important, since the two lines are at very different locations along the echelle order, the first one being very close to the order blue end, and this results in a significant difference reaching 30\% relative variation of the LSF width. We tested the whole procedure for a unique echelle order and for adjacent orders merging. We found that the re-binning and the interpolations required during the order merging as well as the resulting non-monotonous LSF wavelength variation introduce strong residuals at the locations of the deep lines and make the correction more difficult. For this reason we have chosen to restrict the analysis to one order. After final convergence we store the wavelength-dependent LSF width and, importantly, the adjusted transmission model BEFORE its convolution by the LSF.

\subsection{Application, examples of corrections}
\begin{figure}[h]
\centering
\includegraphics[width=1\linewidth]{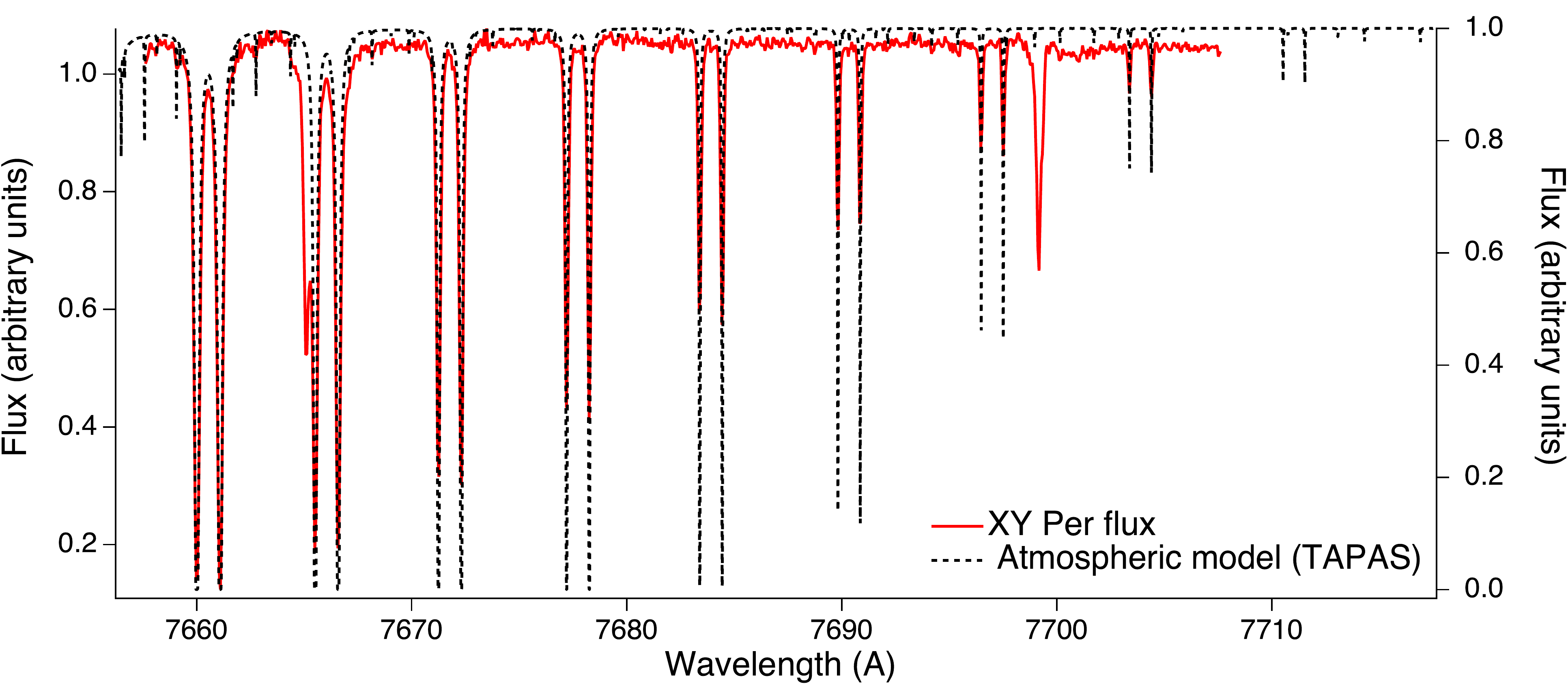}
\includegraphics[width=1\linewidth]{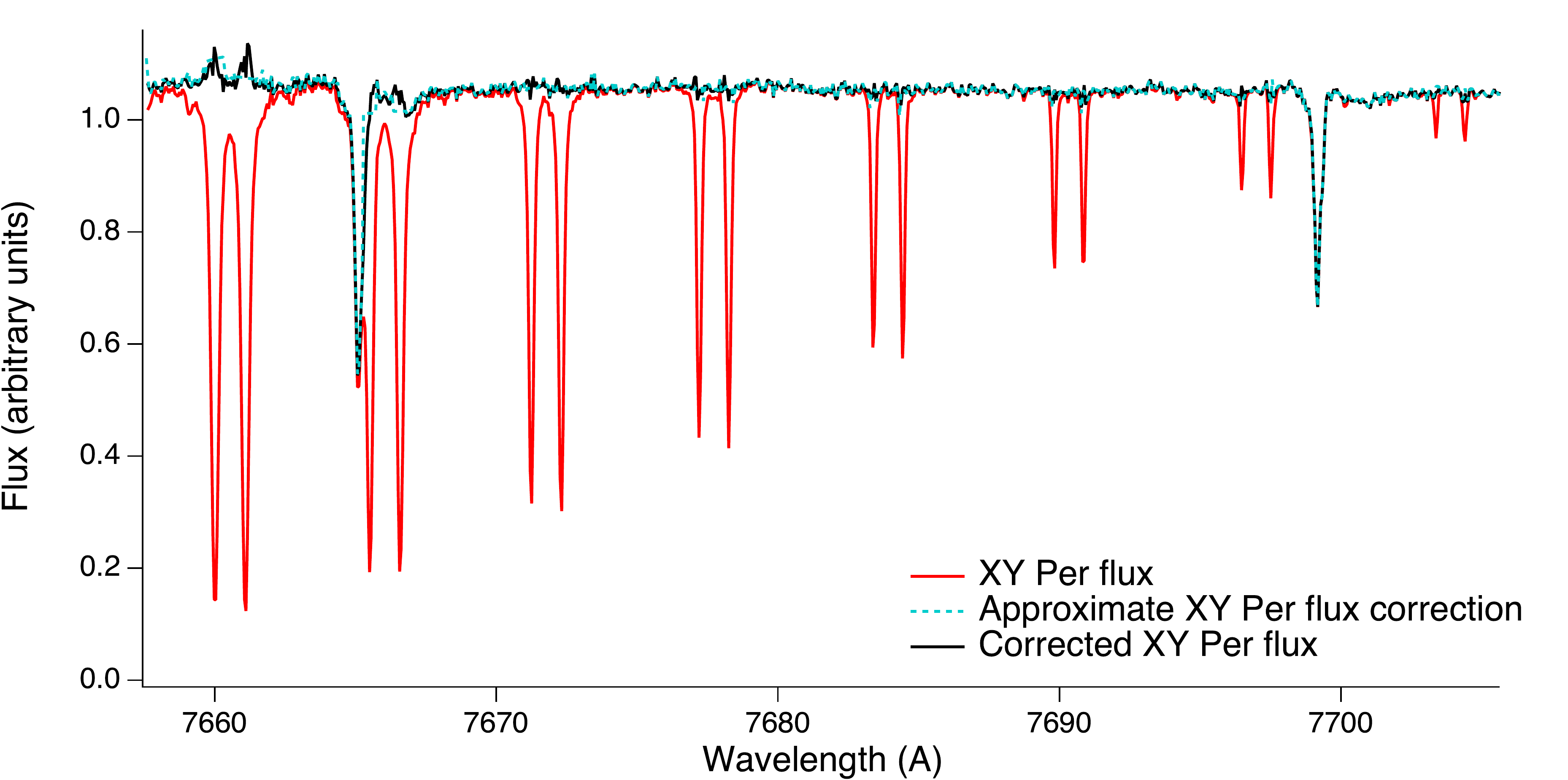}
\caption{Illustration of the telluric correction, prior to the dual interstellar-telluric profile-fitting. The TAPAS model is selected for the Pic du Midi observatory. (a) Atmospheric profile (black color, right axis) after adjustment to the observation, superimposed on the initial stellar spectrum (here the star XY Per, red color, left axis). Note that this is the atmospheric profile before convolution by the instrumental profile, which will be used at the next step of profile-fitting. (b) {\it Corrected} stellar spectrum (black line) obtained by division of the raw data by the above atmospheric profile, after its convolution by the instrumental profile. It is superimposed on the initial stellar spectrum (red line). There are some {\it spiky} residuals, in the strong absorption areas, which are due to the division of  weak quantities for both data and models. Also shown is the {\it quasi-continuum} obtained from the corrected spectrum after some interpolation at these regions  (turquoise dashed line). Here, the residuals are weak and there is only a very small difference between the two curves. Note that what we want to illustrate in this figure are the accuracy of the atmospheric model and the achieved level of adjustment: none of these corrected spectra will enter the dual telluric-interstellar profile-fitting, instead only the adjusted pre-instrumental profile shown at top will be used.}    
\label{fig:tapas}
\end{figure}

This method was applied to all stars, which are listed in Table \ref{tab:targettable} on top of the 2 horizontal lines. A special TAPAS model was used for each separated observatory. 
An example of telluric model adjustment is represented in Figure  \ref{fig:tapas}. The fitted TAPAS transmission before entrance in the instrument is displayed on top of the data in Figure \ref{fig:tapas}(top). At the end of the process, if we convolve this transmission by the LSF and divide data by the resulting convolved model, we obtain a partially corrected spectrum, represented in Figure \ref{fig:tapas}(bottom).
 There are still narrow residuals {\it spikes} at the locations of the strongest lines (see e.g. the first pair of $O_2$ lines in Figure \ref{fig:tapas}(bottom)). This is unavoidable in the case of deep lines since the flux in data and model is close to zero. However, because we will not use the result of the division, but only keep the best telluric model as an ingredient to enter the K\,{\sc i} profile-fitting, we are not impacted by such {\it spikes}, the main advantage of the method. There is an additional advantage: dividing data by a convolved telluric transmission profile is not a fully mathematically correct solution, because the convolution by a LSF of the product of two functions is not equivalent to the product of the two functions separately convolved. The  latter computation gives an approximate solution, however, only if all features in the initial spectrum are  wider than the LSF. This is not the case in the presence of sharp stellar or interstellar features, and this may impact on the accuracy of the results. Note that this type of methods can be used in exo-planetary research for the removal of telluric contamination of the spectra obtained from ground-based facilities.\\

\section{Spectral analysis: interstellar K\,{\sc i} doublet extraction}

\subsection{K\,{\sc i} doublet extraction method}
As a result of the telluric correction procedure, two files are produced for each target, the first one is the fitted TAPAS model, before convolution by the LSF. The second contains the LSF width as a function of wavelength. We use the second to extract the two separate LSFs appropriate for each K\,{\sc i} line (7665 and 7700 \AA . After this preparation, we then proceed to the K\,{\sc i} doublet profile-fitting in a rather classical way \citep[see, e.g.][]{Puspitarini12}, except for two aspects: -(i) the inclusion of the telluric profile, now combined with Voigt profiles before convolution by the LSF and -(ii) the use of two distinct LSFs.\\
The profile-fitting procedure is made of the following steps:
\begin{enumerate}
    \item Conversion of wavelengths into Doppler velocities for both the stellar spectrum and the  fitted transmission model, separately for the two K\,{\sc i} line centers. We restrict to spectral regions wide enough to allow a good definition of the two continua.
    \item Continuum fitting with first, second, or third order polynomial, excluding the interstellar features.  The normalization of the spectra at the two transitions is done through division by the fitted continua. 
    \item Visual evaluation of the minimum number of clouds with distinct Doppler velocities based on the shapes of the K\,{\sc i} lines. Markers are placed at the initial guesses for the cloud velocities, using the 7699 \AA\ line.
    \item  Fitting to normalized data of convolved products of Voigt profiles and the telluric model by means of a classical Levenberg-Marquardt convergence algorithm in both K\,{\sc i} line areas and cloud number increase, if it appears necessary. Here we have imposed for each cloud a maximum Doppler broadening (i.e. combination of thermal broadening and internal motions) of 2.5 km s$^{-1}$. This arbitrary limit is guided by the spectral resolution of the data, the spatial resolution of the present 3D maps of dust, and our limited objectives in the context of this preliminary study (see below).  
\end{enumerate}

As is well known, the advantage of fitting a doublet instead of a single line is that more precise cloud disentangling and velocity measurements can be made, since there must be agreement between shifts, line depths and widths for the two transitions and therefore tighter constraints are obtained. An advantage of our technique is obtaining maximal information from the doublet whatever the configuration of telluric and K\,{\sc i} lines. 
On the other hand, disentangling clumps with velocities closer than $\simeq$ 2 km s$^{-1}$ is beyond the scope of this study, devoted to a first comparison with 3D dust reconstruction. More detailed profile-fitting combining K\,{\sc i} and other species is planned. Accurate measurements of the K\,{\sc i} columns for the various velocity components is similarly beyond the scope of this profile-fitting. Despite the use of the K\,{\sc i} doublet, such measurements are difficult because the lines are often saturated and the widths of the individual components are generally smaller than the LSF for most of the data. WH01 found a median component width (FWHM) lower than 1.2 km s$^{-1}$, compared to average values on the order of 3.5 km s$^{-1}$ for the present data. Here we attempt to use solely the order of magnitude of the measured columns in our search for a counterpart of the absorption in the 3D dust map. 

\subsection{Profile-fitting results}
\begin{figure}
   \centering
   \includegraphics[width=\hsize]{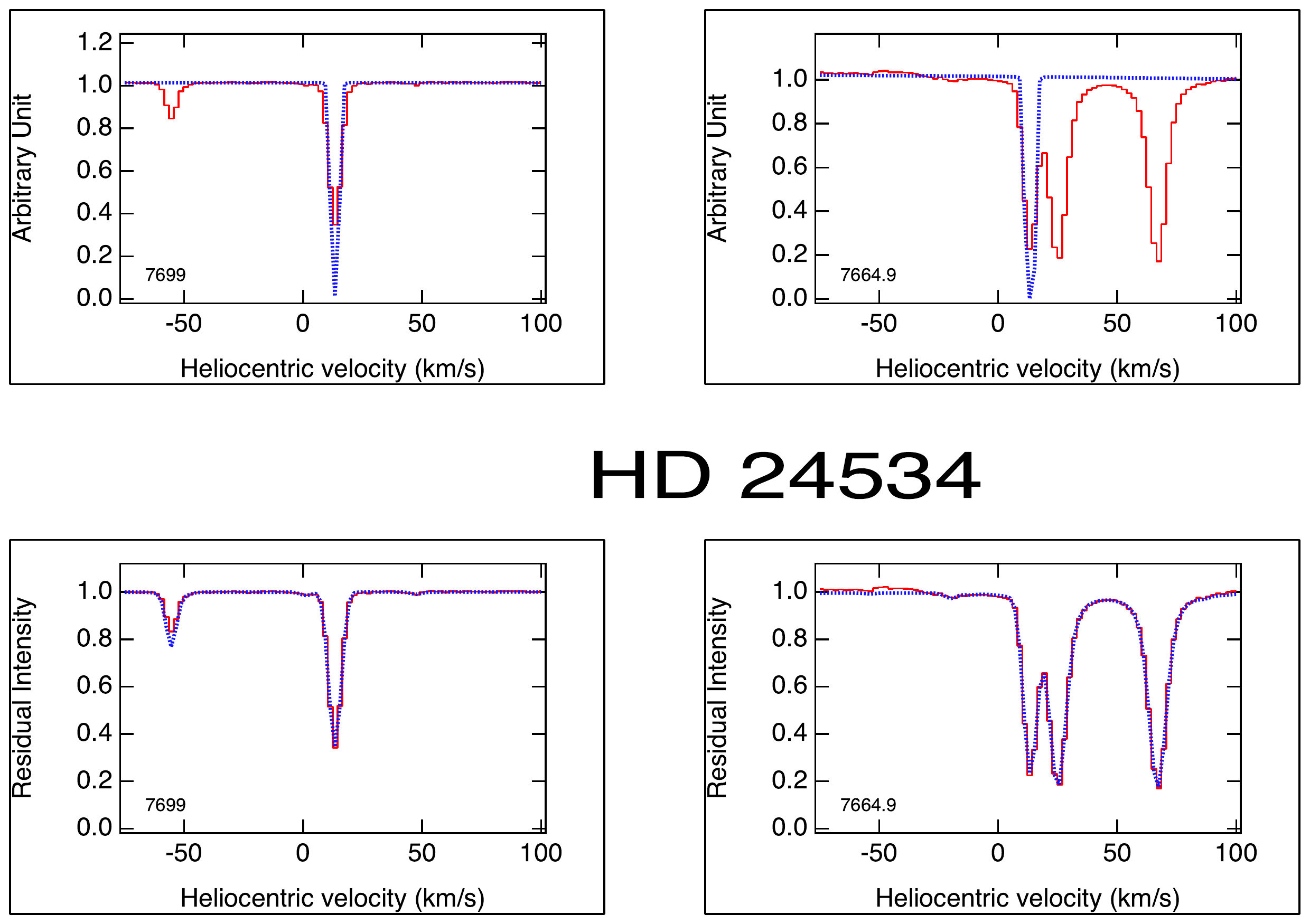}
     \caption{Example of K\,{\sc i} doublet dual interstellar-telluric profile-fitting: star HD 24534 (X Per). The spectra are shown around the 7699 (resp. 7665) \AA\ line at left (resp. at right). The top graphs show the fitted interstellar model before convolution by the instrumental profile (blue line), superimposed on the raw data. The bottom graphs show the adjusted interstellar-telluric model (blue line) superimposed on the normalized data. For this target, the interstellar K\,{\sc i} lines are Doppler shifted out of the strong telluric oxygen lines.
              }
              \label{fig:outO2}
     \end{figure}

\begin{figure}
  \centering
   \includegraphics[width=\hsize]{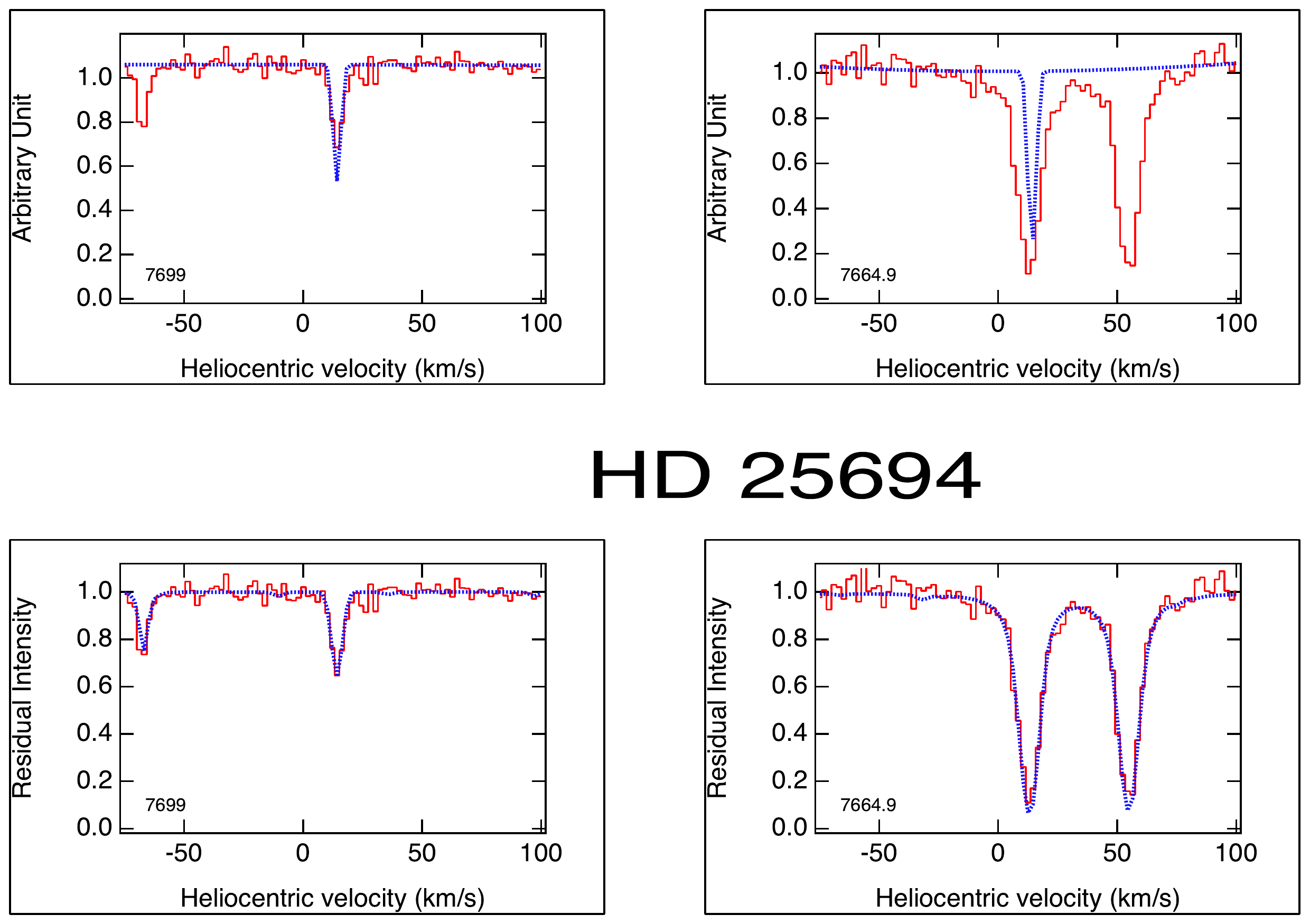}
      \caption{Same as Fig. \ref{fig:outO2}, here for the targets star HD 25694. The 7665 \AA\ K\,{\sc i} line is blended with one of the telluric oxygen lines.
              }
         \label{fig:inO2}
   \end{figure}

\begin{figure}
   \centering
   \includegraphics[width=\hsize]{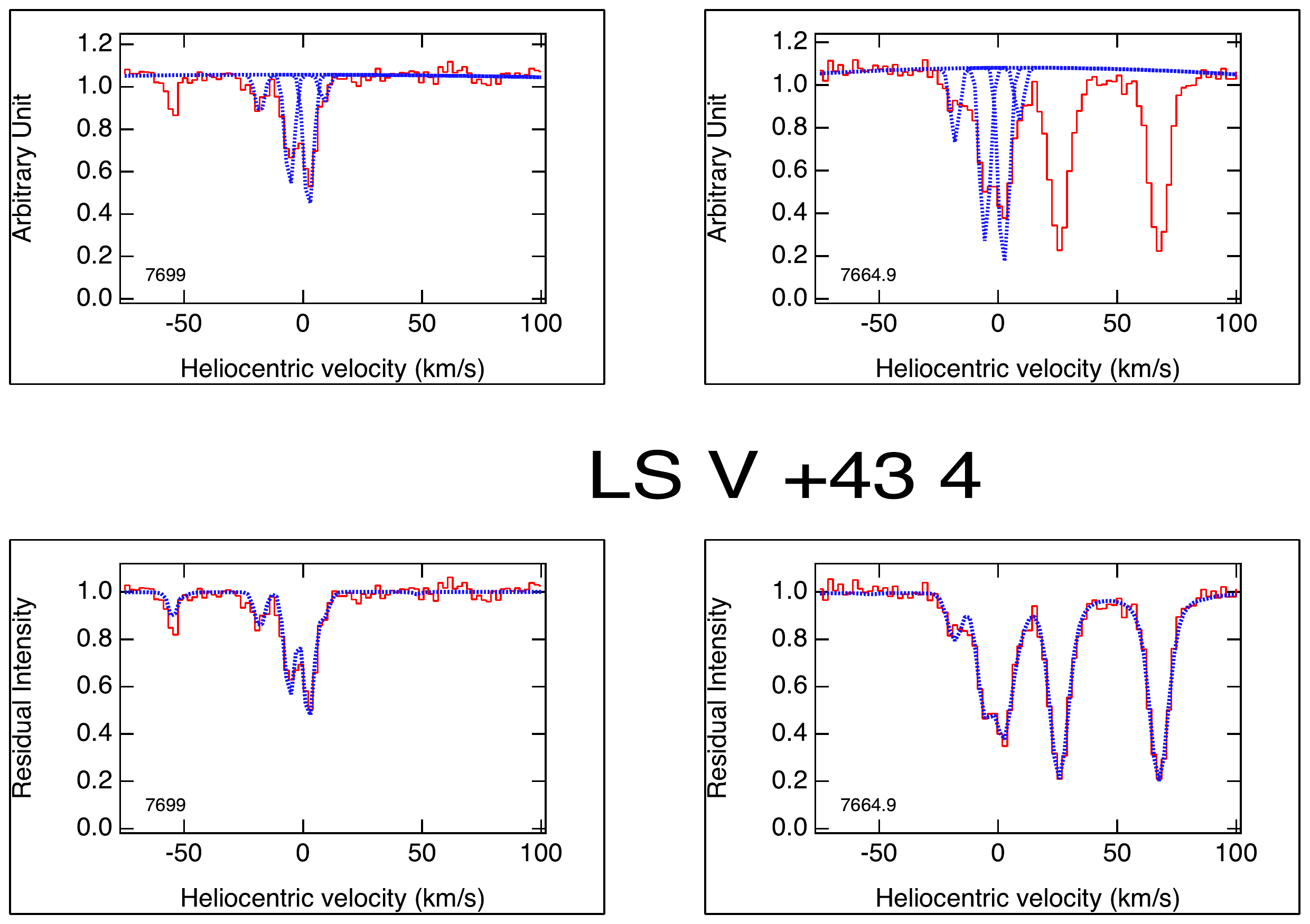}
     \caption{Same as Fig. \ref{fig:outO2}, here for the target star LS V +43 4. 4 individual clouds are necessary to obtain a good fit, i.e with residuals within the noise amplitude
              }
              \label{fig:4clouds}
     \end{figure}

\begin{figure}
      
   \centering
   \includegraphics[width=\hsize]{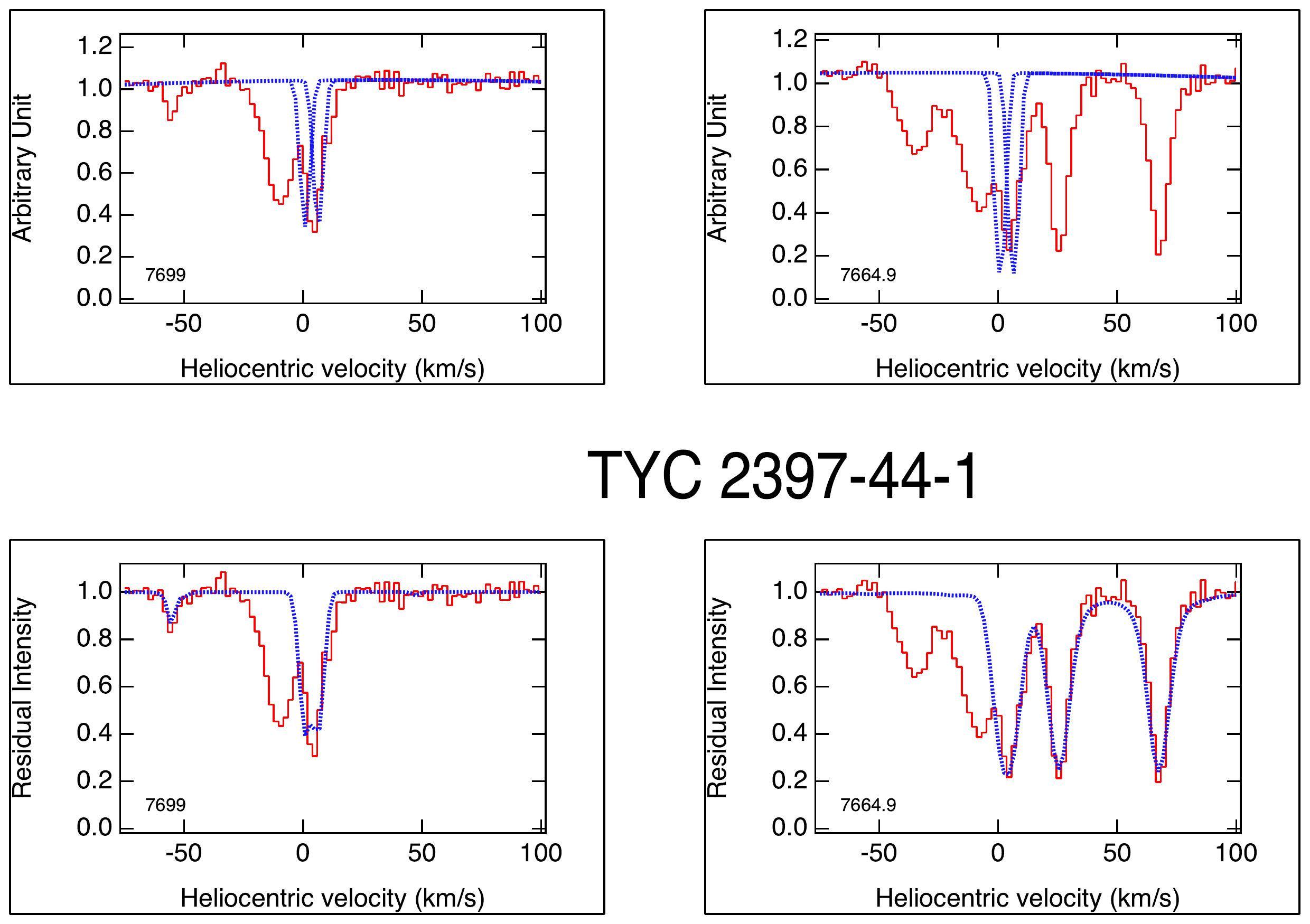}
     \caption{Same as Fig. \ref{fig:outO2}, for the star TYC-2397-44-1, as an example of profile-fitting in presence of strong, but Doppler shifted stellar lines
              }
             \label{fig:stellar}
  \end{figure}

Examples of profile-fitting results are presented in figures 3 to 6. We have selected different situations in terms of locations of the interstellar K\,{\sc i} lines with respect to the strong telluric oxygen lines and in terms of cloud numbers. Fig.\ref{fig:outO2} shows a case in which both K\,{\sc i} lines are shifted out of the $O_2$ telluric lines and there is only one detected strong velocity component. Fig.\ref{fig:inO2} displays a case in which the K\,{\sc i} 7664.9 absorption area coincides with the center of a strong $O_2$ telluric line. Still, the inclusion of an interstellar absorption contribution in this area is required to obtain a good adjustment and there is some information gathered from using the doublet instead of the K\,{\sc i} 7699 line solely. Fig. \ref{fig:4clouds} corresponds to 4 distinct absorption velocities. Finally, Fig. \ref{fig:stellar} corresponds to a special case of a star cooler than the majority of other targets and possessing strong K\,{\sc i} lines, here fortunately sufficiently Doppler shifted to allow measurements of the moderately shifted Taurus absorption lines.
\\
Profile-fitting results are listed in Table \ref{tab:velocitytable}. Note that errors on velocities and K\,{\sc i} columns have a limited significance and must be considered cautiously. Due to the partially arbitrary choice of cloud decomposition, our quoted errors on velocities correspond to this imposed choice. It has been demonstrated that there is a hierarchical structure of the dense, clumpy ISM in regions such like Taurus, and that a decomposition into more numerous sub-structures with relative velocity shifts on the order of 1 km s$^{-1}$ is necessary if the spectral resolution is high enough \citep[][,WH01]{Welty94}. However, as we will see in the next section, the 3D maps presently available do not allow to disentangle the corresponding clumps that have a few parsec or sub-parsec sizes and this justifies the choice of our limitations during the profile-fitting. As a consequence of our allowed line broadening, uncertainties on the central velocities of the clouds may reach up to $\simeq$ 2 km s$^{-1}$. We have checked our fitted radial velocities with those measured by  WH01 for the targets in common, $\zeta$Per, $\xi$Per and $\epsilon$Per, and found agreement within the uncertainty quoted above, and taking into account our coarser velocity decomposition. On the other hand,  our quoted errors on the K\,{\sc i} columns also correspond to our choices of decomposition and are lower limits. The order of magnitude of the columns is a precious indication for the following step of velocity assignment to structures revealed by the maps. 


\section{Updated 3D dust maps}
\subsection{Map construction}
Massive amounts of extinction measurements towards stars distributed in distance and direction can be inverted to provide the location, in 3D space, of the masses of interstellar dust responsible for the observed extinction. Several methods have been used and the construction of 3D dust maps is in constant progress, due the new massive stellar surveys, and, especially, due to parallaxes and photometric data from the ESA Gaia mission. Here we use an updated version of the 3D extinction map presented by \cite{Lallement19}, that was based on a hierarchical inversion of extinctions from Gaia DR2 and 2MASS photometric measurements on the one hand, and Gaia  DR2 parallaxes on the other hand. The inversion technique here is the same as for this previous map, however, the inverted extinction database and the prior distribution are different. In the case of the construction of the previous map, a homogeneous, plane-parallel distribution was used as the prior. Here instead, the prior distribution is the previous 3D map itself. The inverted dataset is no longer the set of Gaia-2MASS extinctions, instead it is now the extinction database made publicly available by \cite{Sanders18}, slightly augmented by the compilation of nearby star extinctions used in \cite{Lallement14}. With the goal of deriving their ages for a large population of stars, \cite{Sanders18} have analyzed, in a homogeneous way, data from six stellar spectroscopic surveys, APOGEE, GALAH, GES, LAMOST, RAVE, and SEGUE, and have estimated the extinction for a series of 3.3 million targets (if the catalog is restricted to the "best" flag, see \cite{Sanders18}). To do so, and in combination with the parameters inferred from the spectroscopic data, they used the photometric measurements from various surveys and in a large number of bands (J, H, K from 2MASS, G, GBP, GRP from Gaia, gP, rP, iP from Pan-STARRS, g, r, i from SDSS, Jv ,Hv, Kv from VISTA). They used a Bayesian model and priors on the stellar distributions as well as priors on the extinction derived from the Pan-STARRS 3D mapping \citep{Green18} or the \cite{Drimmel03} large-scale model. Because the extinction determined by \cite{Sanders18}  is in magnitude per parsec in the V band, and our prior map was computed as A$_{0}$ (i.e. the mono-chromatic extinction at 5500 \AA~), we applied a fixed correction factor to the prior distribution of differential extinction, a factor deduced from cross-matching A$_{0}$ extinctions used for this prior and the \cite{Sanders18} extinctions for targets in common. Due to the strong constraints brought by the spectroscopic information, the individual extinctions from the \cite{Sanders18} catalog are more accurate than purely photometric estimates, and are expected to allow refining the 3D mapping. This is why the minimal spatial correlation kernel for this new inversion, a quantity that corresponds to the last iteration of the hierarchical scheme, is 10 pc, compared to the 25 pc kernel of the Gaia-2MASS map (see \cite{Vergely10} for details on the basic principles and limitations of the inversion, and \cite{Lallement19} for a description of the hierarchical inversion scheme). Note, however, that, because the whole set of spectroscopic surveys is far from covering homogeneously all directions and distances, such a spatial resolution is not achievable everywhere and, in the case of regions of space that are not covered, the computed solution remains the one of the prior 3D map.

\begin{figure}
\centering
 \includegraphics[width=\hsize]{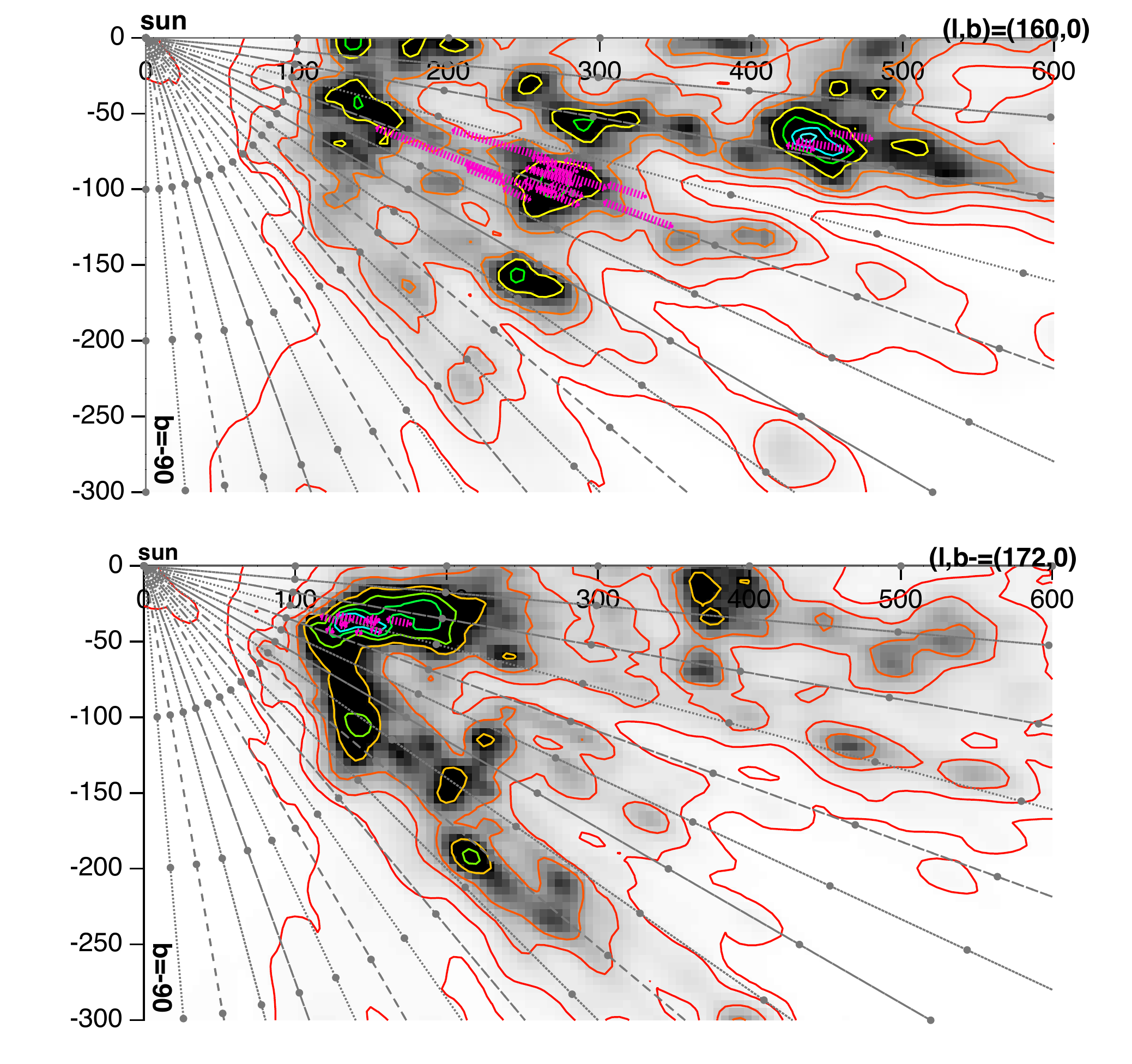}
 \caption{Images of the inverted dust distribution in two vertical planes containing the Sun (located at 0,0) and oriented along  Galactic longitudes l=160 and 172 degrees. Galactic latitudes are indicated by dashed black lines, every 5 degrees from b=0 in the Plane to b=-90 (South Galactic pole). Coordinates are in parsecs. The quantity represented in black and white is the differential extinction dA$_{V}$/dl at each point. Isocontours are shown for 0.0002,0.00045, 0.001, 0.002, 0.005, 0.01 , 0.015 and 0.02 mag per parsec. Superimposed are the closest and farthest locations of the molecular clouds from the \cite{Zucker20} catalog that are at longitudes close to the one of the image (see text). A dashed pink line connects these two locations for each cloud.}
\label{fig:examp_3D_compzucker}
\end{figure}

\subsection{Taurus dust clouds in 3D}

The Taurus area has been quite well covered by the spectroscopic surveys used by \cite{Sanders18}, in particular APOGEE and LAMOST \citep{Majewski17,Deng12}. This has allowed a more detailed mapping of the dust clouds in this region and justifies the use of the updated map for our kinetic study. In Figure \ref{fig:examp_3D_compzucker} we show images of the newly computed differential extinction, in two planes perpendicular to the mid-Plane and containing the Sun, oriented along Galactic longitudes 160 and 172\fdeg respectively. The color-coded quantity is the local differential extinction, or extinction per distance, a quantity highly correlated with the dust volume density. As mentioned above, an important characteristic of the maps and images is the minimum size of the cloud structures, directly connected to the minimum spatial correlation kernel imposed during the inversion. In the Taurus area, thanks to the target density the kernel is on the order of 10 pc almost everywhere in the regions of interest up to $\simeq$ 500 pc.
We have compared the locations of the dense structures reconstructed in Taurus, Perseus and California with results on individual molecular clouds distances in the recent catalog of \cite{Zucker20}, based on PanSTARRS and 2MASS photometric data on the one hand, and Gaia DR2 parallaxes on the other hand. Superimposed on the first image along l=160\fdeg   (resp. 172\fdeg) are the closest and farthest locations of the clouds as measured by \cite{Zucker20}, restricting to cloud centers with longitudes comprised between 156 and 164\fdeg (resp. 166 and 176\fdeg). This implies that we neglect here the difference of up to 4\fdeg between the molecular cloud actual longitudes and the longitude of the vertical plane. We also used solely the statistical uncertainties on the distances quoted by  \cite{Zucker20}, and did not add the systematic uncertainty of 5 $\%$ quoted by  the authors. {Despite these differences and reduced uncertainties, and despite our limited spatial resolution required by the full-3D inversion technique, it can be seen in Figure \ref{fig:examp_3D_compzucker} that there is a good agreement between the locations of the dense structures obtained by inversion and the locations independently obtained by \cite{Zucker20}.}  

As mentioned in the introduction, large amounts of YSOs can now be identified and located in 3D space, and several studies have been devoted to their clustering and associations with known molecular clouds \citep{Grossschedl21, Roccatagliata20, Kounkel18,Galli19}. If the stars are young enough, they remain located within their parent cloud and their proper motions are identical to the motion of the cloud. As an additional validation of the new 3D dust distribution, we have superimposed the locations of the Main Taurus YSO clusters found by \cite{Galli19} in the dust density images used for the preliminary 3D kinetic tomography presented in the next section.  The locations of these YSO clusters are also, in principle, those of the associated molecular clouds. L 1517/1519, L 1544, L 1495 NW, L 1495, B 213/216, B 215, Heiles Cloud 2, Heiles Cloud 2NW, L 1535/1529/1531/1524, L 1536, T Tau cloud, L 1551 and L 1558 are the main groups identified by the authors and are shown in figures \ref{fig:lon1},\ref{fig:lon2},\ref{fig:lon3}. Although the dust map spatial resolution is not sufficient to disentangle close-by clouds, it can be seen from the series of figures that the YSO clusters coincide with very dense areas.

\subsection{Link between extinction and K\,{\sc i} absorption strength}

\begin{figure}
\centering
 \includegraphics[width=\hsize]{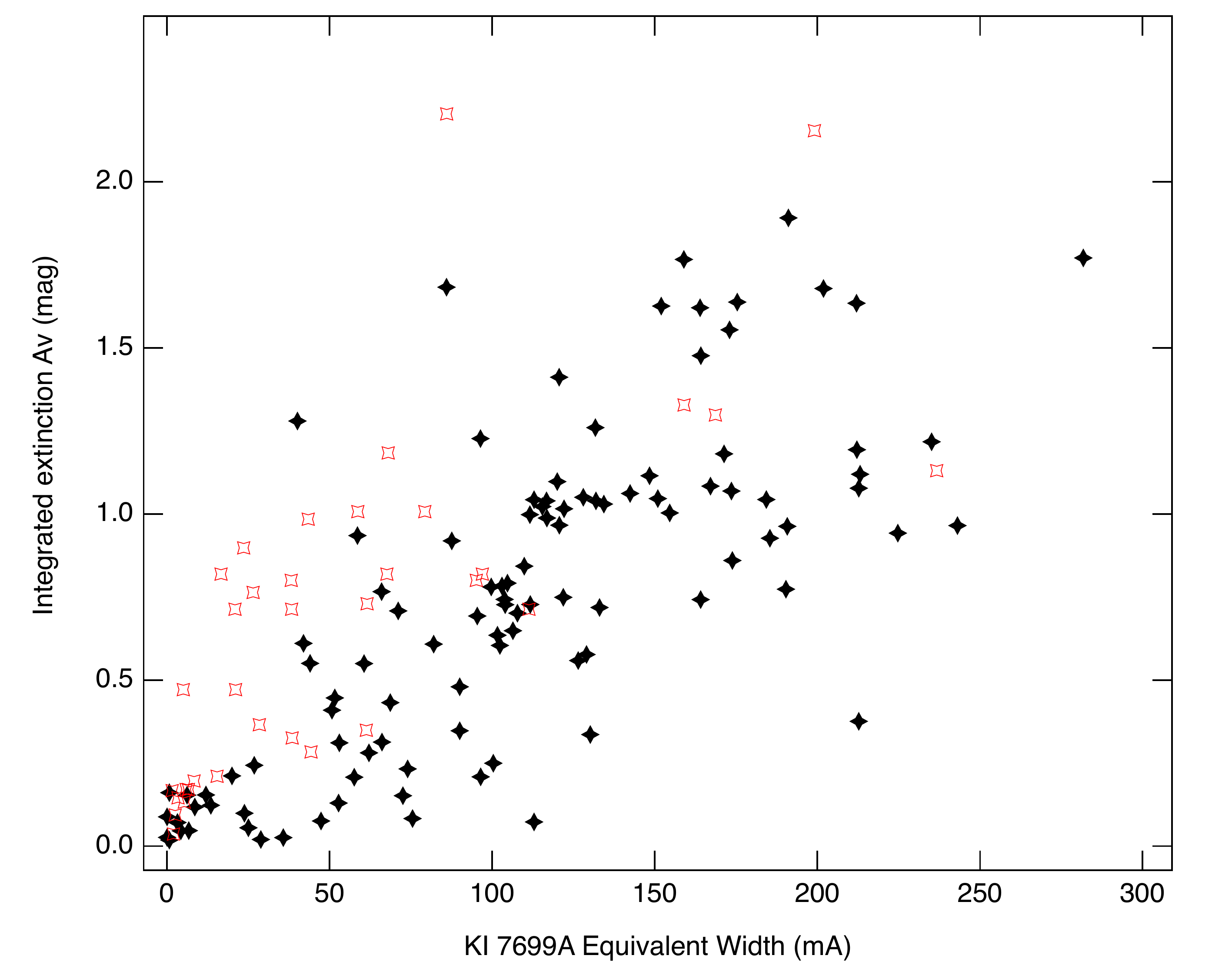}
 \caption{Integrated extinction between the Sun and each target, using the 3D map, as a function of the full equivalent width of the  7699\AA~ K\,{\sc i} absorption profile (black signs). Also shown are 7699\AA~ K\,{\sc i}  equivalent widths from WH01  and extinctions estimated from conversion of H columns (red signs, see text).}
\label{Av_KI_relax}
\end{figure}

Most of our targets do not have individually estimated extinctions. However, it is possible to use the 3D distribution of differential extinction and to integrate this differential extinction along the path between the Sun and each target star, to obtain an estimate of the star extinction. We have performed this exercise for our series of targets and have compared the resulting extinctions A$_{V}$  with the full equivalent widths of the 7699\AA~ K\,{\sc i} absorption profiles. The comparison is shown in Figure \ref{Av_KI_relax}. Despite an important scatter, there is a clear correlation between the two quantities, a confirmation of the link between neutral potassium and dust, and a counterpart of the link between K\,{\sc i} and the H column shown by WH01. Note that we do not expect a tight correlation between the two quantities for several reasons. One of the reasons is the limited resolution of the map, and the fact that the extinction is distributed in volumes on the order of the kernel. As a consequence, high density regions in cloud cores are smeared out and high columns that occur for lines of sight crossing such cores have no comparably high counterparts in the integrations. A second reason is the distance uncertainty. Despite their unprecedented accuracy, Gaia EDR3 uncertainties are on the order of a few pc  for the Taurus targets and this may influence the integral. An additional source of variability of a different type is the non linear relationship between the equivalent width we are using here and the column of atoms,  and its  dependence on the temperature, turbulence and velocity structure (see, e.g., WH01).  In order to compare the links between K\,{\sc i} and H columns one the one hand, and between K\,{\sc i} and dust columns on the other hand, we have used the 7699\AA~ K\,{\sc i} equivalent widths measured by WH01 and we have converted  H$_{tot}$ columns taken from  the same study into extinctions, assuming a constant ratio and A$_{V}$=1 for  N(H$_{tot}$)= 6. 10$^{21}$/3.1 cm$^{-2}$.  Figure \ref{Av_KI_relax} displays the relationship between the two quantities, superimposed on our values of equivalent widths and extinctions. The scatter appears similar, as confirmed by the similarity of the Pearson correlation coefficients, 0.74 for our results and  0.72 for the quantities derived from the WH01 study. This suggests that, as expected, the link of K\,{\sc i} with dust  is comparable to its link with H. Note that our 105 data points correspond to Taurus stars, while the 35 ratios using K\,{\sc i} and N(H$_{tot}$) from WH01 correspond to stars distributed over the sky. The average ratio between extinction and K\,{\sc i}  is smaller in our case, a large part of the difference is certainly linked to our limited spatial resolution and the corresponding smearing of the extinction. In the context of kinetic tomography, the obtained relationship is encouraging, and suggests that K\,{\sc i} absorptions may be a convenient tool. 

\section{Preliminary 3D kinetic study}

Our objective here is to study which constraints brought by our relatively small number of K\,{\sc i} measurements in the context of kinetic tomography, more precisely what velocity assignments can be made to dense structures reconstructed in 3D by extinction inversion, using exclusively the measured absorption velocities and our most recent 3D dust maps. We started by extracting from the 3D distribution of differential extinction a series of images of dust clouds in vertical planes containing the Sun and oriented towards Galactic longitudes distributed between  l=150 and l= 182\fdeg, by steps of 2.5\fdeg, at the exception of l=152.5\fdeg, due to the absence of targets, i.e. images similar to those in figure \ref{fig:examp_3D_compzucker}, now covering the whole Taurus-Perseus-California regions. About these images, it is important to keep in mind that the mapped structures have a minimum size of 10 pc, which prevents detecting smaller clumps. We have used this set of images, the paths to the target stars and the absorption results to link in a non-automated way dust concentrations and velocities, plane after plane. To do this, for each image of the vertical plane, we selected the target stars whose longitudes are less than 1.3 degrees from the longitude of the plane and we plotted the projection of their line of sight on the plane. We did this for their most probable distance, in all cases available from the Gaia Early Data Release 3 (EDR3) catalog \citep{Brown20} except for two targets. We then compared their observed K\,{\sc i} absorption velocities as well as their associated approximate columns (in units of $10^{10}$ cm$^{-2}$) and the dense structures encountered along their trajectories.  As we have already pointed out, this exercise would take too much time in the case of more numerous data and the results are partially arbitrary, but we want to illustrate here what can be done based only on a preliminary visual method. Work is underway to develop automated techniques. 

{Fig. \ref{fig:l150} displays the vertical plane along l=150\fdeg. Planes along other longitudes are displayed in Figures \ref{fig:lon1},\ref{fig:lon2},\ref{fig:lon3}. On each figure we restrict to radial velocities measured for stars with longitudes within 1.3\fdeg from the plane, listed at bottom. To help visualizing which star corresponds to which velocity assignment, we have numbered them and the numbers correspond to the objects listed in the text included in each figure. The velocity assignments to the dense clouds are indicated by arrows pointing outwards (inwards) for positive (negative) radial velocities. The length of the arrow is proportional to the velocity modulus. Two or more velocities at very close locations in the map indicate that information on the same cloud is provided by different stars. In some cases the arrows are slightly displaced to avoid superimposition of different measurements based on stars at very close latitudes. In all cases, differences between measured velocities on the order of 1 km s$^{-1}$ and sometimes up to 2 km s$^{-1}$  may be due to profile-fitting uncertainties and do not necessarily imply a different clump. 

The first result is linked to the exercise itself: it was surprisingly easy to associate velocities and clouds. For all K\,{\sc i} columns on the order or above $\simeq$ 5 10$^{10}$ cm$^{-2}$, dense structures are found along the corresponding path to the target star and the cases with multiple velocities offered often obvious solutions. In general, K\,{\sc i} starts to get detected for lines-of-sight that are crossing the 0.005 mag pc$^{-1}$ differential extinction iso-contour, although this approximate threshold  varies as a function of the distance to the Plane and the signal to noise ratio of the recorded spectrum. There is consistency between the near side of the reconstructed main Taurus dust clouds and the distances to the targets with detected absorption. The closest star with detected K\,{\sc i} is HD26873 at 123.5$\pm$1.3 pc, followed by HD25204 at 124$\pm$6.8 pc, and DG Tau at 125.3$\pm$ 2 pc. Only HD23850 at 123.2$\pm$ 7.3 pc has a smaller parallax distance, but the uncertainty on its distance is significant. Similarly, HD280026 has no detected neutral potassium, but deep 5890 \AA~ sodium absorption despite a most probable distance as short as 103 pc. However, there is no DR2 nor EDR3 Gaia parallax value, and this unique distance estimate is based on Hipparcos, with a large associated uncertainty of 30 pc. As a conclusion, it is likely that these last two targets are located within the near side boundary of the Taurus complex, probably in the atomic phase envelop and very close to the molecular phase at $\simeq$ 123 pc. 
 These are encouraging results, showing that the 3D maps, the distances, and the velocity measurements are now reaching a quality that is actually allowing kinetic tomography. In many cases the assignment is not questionable (see, e.g. the star 55 in the l=165\fdeg image). In other numerous cases, the assignment based on one star is strongly influencing the solution for other stars. Finally, since there is continuity in the velocity pattern in consecutive planes, several consecutive planes can be used to follow both the shape and velocity changes of the same large structure. 
 
 We now discuss the results in more detail, plane by plane. The l=150\fdeg image shows two dense structures at 145 and 400 pc and a weaker cloud at 250 pc. Interestingly, the three distances are similar to the distances to Main Taurus, California and Perseus, but these opaque regions are located much closer to the Galactic plane than the well known dense clouds. As can be seen from all figures (see, e.g., Fig. \ref {fig:lon1}), this triple structure is still present at higher longitudes up to l=170\fdeg. In the l=150\fdeg plane, at these three groups is clearly associated a radial velocity decrease with distance, from $\simeq$+5 down to $\simeq$-8 km s$^{-1}$, i.e. there is a compression of the whole region along the radial direction.  The intermediate structure at 250pc has a very small or null LSR radial velocity. The l=155\fdeg image confirms the velocity pattern (despite the small number of stars) and starts to show the densest cloud complexes, better represented in the next vertical planes. Perseus reaches maximal density at  l$\simeq$157.5, and at this longitude is connected to a series of structures located between 250 and 320 pc and reaching the Galactic Plane. California is the most conspicuous structure at l$\simeq$160 to 162 \fdeg, and starts to be divided in two main parts at l$\simeq$162 \fdeg, the most distant one reaching 540pc. Along l=157.5\fdeg, the global velocity pattern is slightly changing, with appearance of a positive velocity gradient between Main taurus and Perseus. The structure at 320 pc (above Perseus in the image) is peculiar, with a larger positive radial velocity (marked by a question mark), but further measurements are required to conclude. We have located in the l=157.5\fdeg and l=160\fdeg images the dense molecular clouds L1448, L1451 and IC348 in Perseus at their distances found by \cite{Zucker18} and we have compared the radial velocities we can assign to L1448 and IC348 with the radial velocities deduced by the authors based on CO data and 3D extinction. From the strong absorption towards BD+30 540 (star 48) and unambiguously assigned to the region with maximal opacity that corresponds to L1448, we derive V$_{r}$= +5 km s$^{-1}$ for this cloud, close to the 4.8 km s$^{-1}$ peak-reddening radial velocity of \cite{Zucker18}. From several targets crossing the IC348 area (see the next l=160\fdeg plane) we derive V$_{r}$ $\simeq$ +8 km s$^{-1}$ for IC348, close to their average velocity of +8.5 km s$^{-1}$. Still in the l=160\fdeg plane, the structure in the foreground of IC348 at $\simeq$ 140 pc (a distance in agreement with \cite{Zucker18}, see their Fig. 6) appears to have a complex velocity structure with smaller velocities for the dense central region. However, again more data is needed to better map and quantify this property. California at 450 pc is characterized by a decreasing modulus of the inwards motion with increasing distance from the Galactic plane.
 
In the l=165\fdeg  vertical plane one starts to see the disappearance of California and Perseus, and the change of California radial velocities, from negative to null values at about 480 pc. In this image and all the following ones (up to l= 182.5\fdg) the main, densest areas of Main Taurus are the most interesting structures. The YSO clusters derived by \cite{Galli19} in Taurus are found to be located in the densest parts, and the maps confirm the distribution in radial distance of the densest areas found by the authors and previous works, from 130 to 160 pc in the case of all clouds at latitudes below the b$\simeq$-7\fdeg area. An interesting result is the existence of a series of measurements of small radial velocities we could associate with the most distant parts of the clouds (see the four images in Fig \ref{fig:lon2}). We can not distinguish these velocities from the main flow in all stars, only excellent signal to noise and spectral resolution allow to do it, however, their numbers strongly suggests the existence of a negative gradient directed outwards, and this compression is very likely connected to the large number of star-forming regions in this area. The case of L1558 is peculiar. It is the only well known structure found to be locate in a moderately dense area, however this may be an imperfection of the map due the lack of stars because it is located behind a particularly opaque area.  From the point of view of the kinematics, we have used the \cite{Galli19}, positions and velocities in cartesian coordinates to compute the radial velocities in heliocentric and LSR frame, and compared the results with our assignments, when possible. The results  are found to be compatible within 1.5 km s$^{-1}$. For L1495 (l=170\fdeg figure) the most appropriate measurement is for star 74, with a velocity of +7.4 km s$^{-1}$ for the main component, the closest to the Sun. According to \cite{Galli19}, the radial velocity of the cluster  is +6.9 km s$^{-1}$. For B215 we can use several stars that provide velocities +8, +8.5, +7.2, +5.8, +6.5 km s$^{-1}$, i.e. an average of $\simeq$+7km s$^{-1}$, while \cite{Galli19} estimated value is +7.5 km s$^{-1}$. For L1517-1519 we can use star 9 with +6.3 km s$^{-1}$, while the authors find +4.8 km s$^{-1}$. For L1536, our low velocity value from star 65 is +4.9 km s$^{-1}$, compatible with +5.9 km s$^{-1}$. For T Tau, we can use the target star 23 which has two components at +7 and+12 km s$^{-1}$. The cluster found by \cite{Galli19} is apparently located on the more distant, low velocity side with a value of +7.8 km s$^{-1}$. About L1544, the star 44 has its high velocity component at +6.3 km s$^{-1}$, compatible with the +7.6 km s$^{-1}$ found for the cluster. For all these cases, the comparisons seem to confirm the negative velocity gradient we discussed above, i.e., a general compression and deceleration along the radial direction. L1551 does nit follow the rule and there is no agreement between our estimated velocity pattern and the \cite{Galli19} result: if the cluster and associated cloud is located as shown in the l=180\fdeg plane, it should be within the high velocity part of the two regions probed by star 7, i.e. at +8.4 km s$^{-1}$, however, its velocity of +5.6 km s$^{-1}$ is closer to the low velocity component detected for this star, +4.1km s$^{-1}$. On the other hand, in this very thick region the velocity pattern may be more complex and require more constraints.

\begin{figure*}
\centering
  \includegraphics[width=\hsize]{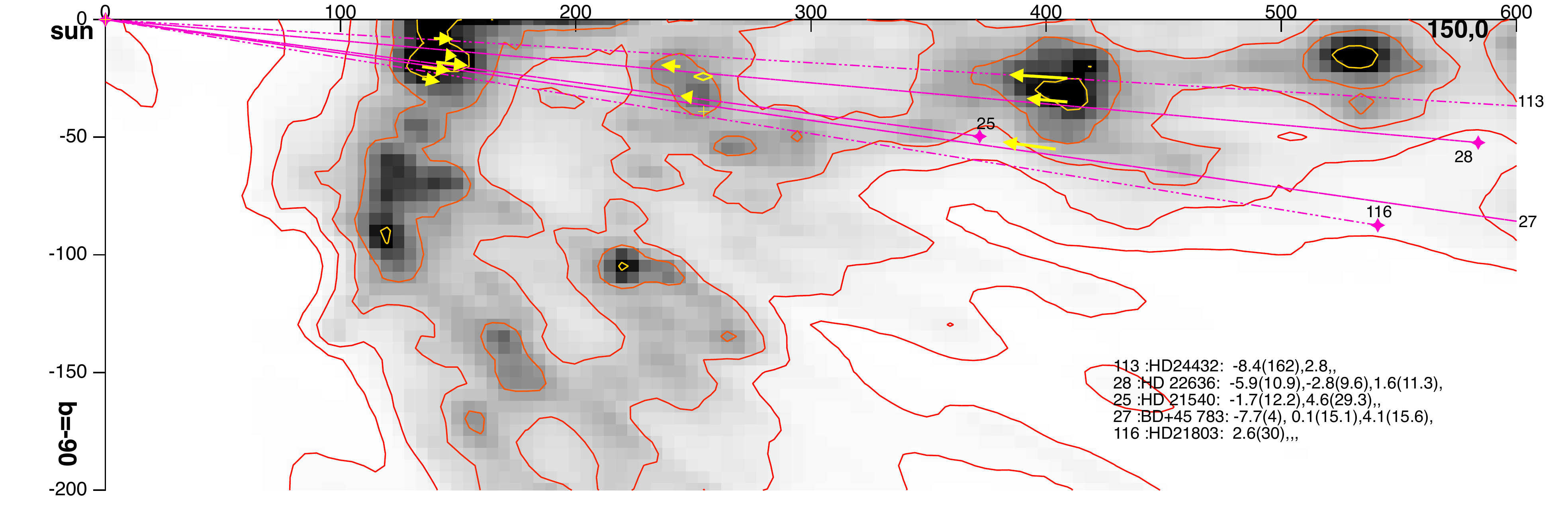}
 \caption{Image of the inverted dust distribution in a vertical plane containing the Sun (located at 0,0) and oriented along  Galactic longitude l=150\fdeg. Coordinates are in parsecs. The quantity represented in black and white is the differential extinction. Iso-contours are shown for 0.0002,0.00045, 0.001, 0.002, 0.005, 0.01 , 0.015 and 0.02 mag per parsec. Superimposed are the paths to the target stars whose longitudes are within 1.3 degrees from the longitude of the represented plane. The stars are numbered  as in the text drawn at bottom in the figure and the Doppler velocities and approximate columns of absorbing K\,{\sc i} in 10$^{10}$ cm$^{-2}$ units are listed for each target (in parentheses after velocities). Stars numbers are indicated in yellow or black. The list of targets is given by decreasing latitude. For distant targets falling outside the image, their numbers are indicated along their paths at boundary of the figure. The preliminary velocity assignments to the dense clouds are indicated by arrows pointing outwards (inwards) for positive (negative) radial velocities. The length of the arrow is proportional to the velocity modulus.}
\label{fig:l150}
\end{figure*}

\begin{figure*}
\centering
  \includegraphics[width=0.9\hsize]{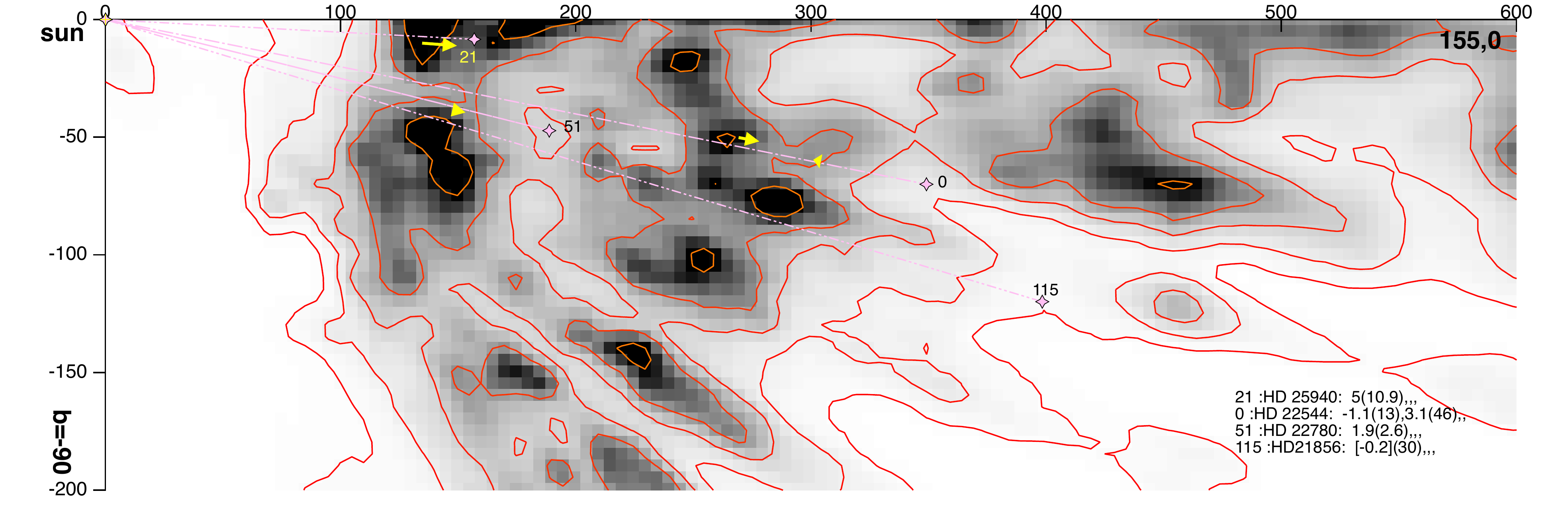}
  \includegraphics[width=0.9\hsize]{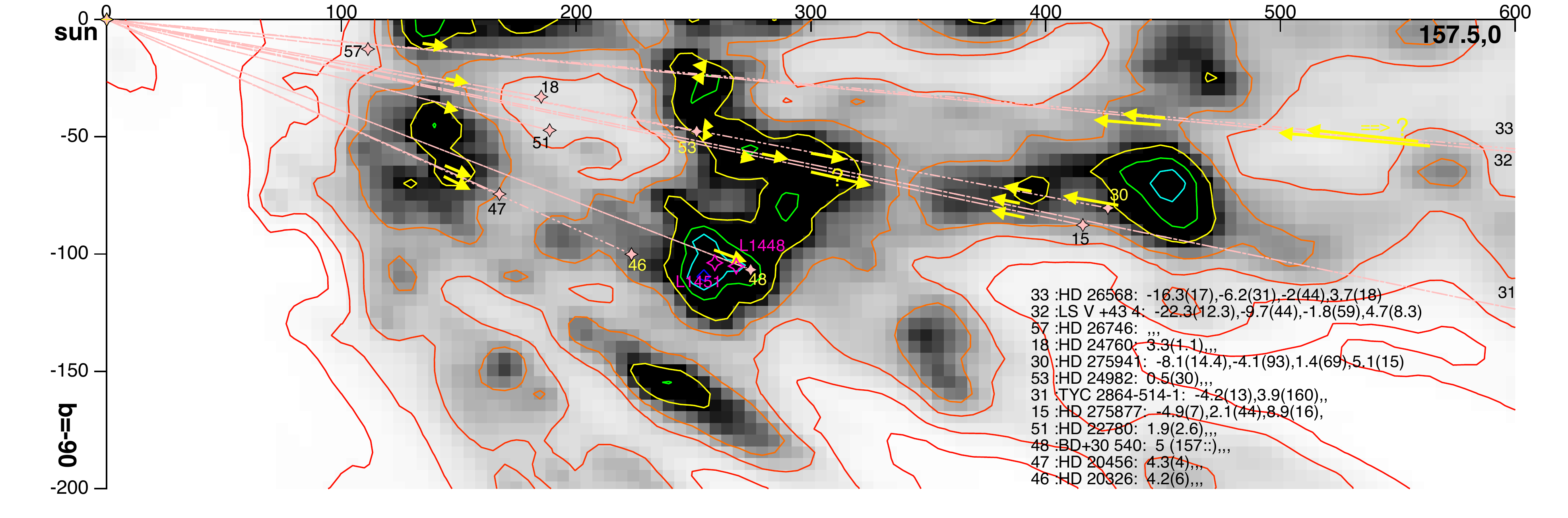}
  \includegraphics[width=0.9\hsize]{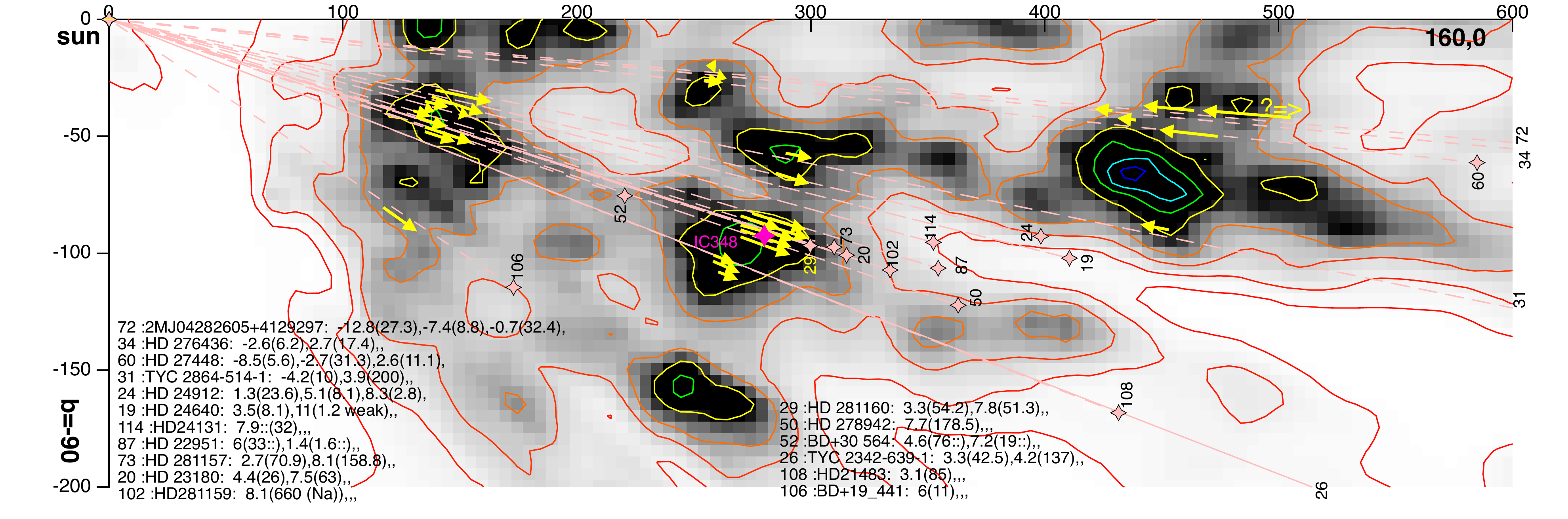}
  \includegraphics[width=0.9\hsize]{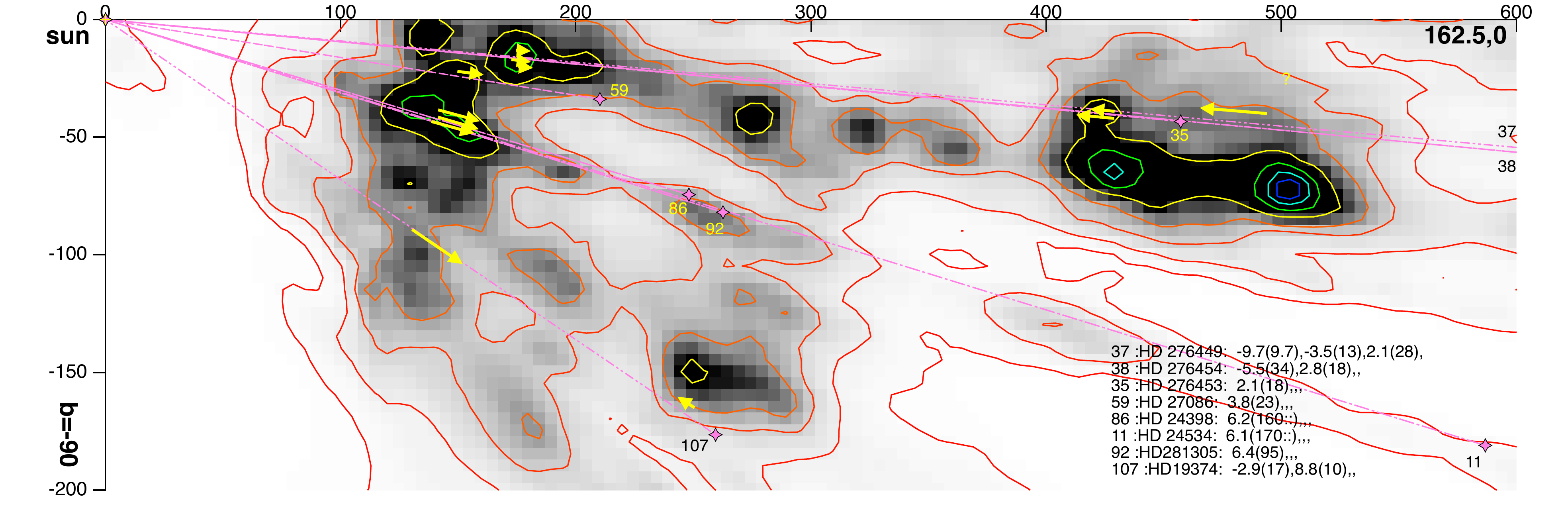}
\caption{Same as Fig \ref{fig:l150}, for longitudes l=155, 157.5, 160 and 162.5\fdeg}
\label{fig:lon1}
\end{figure*}
  
\begin{figure*}
\centering
  \includegraphics[width=0.9\hsize]{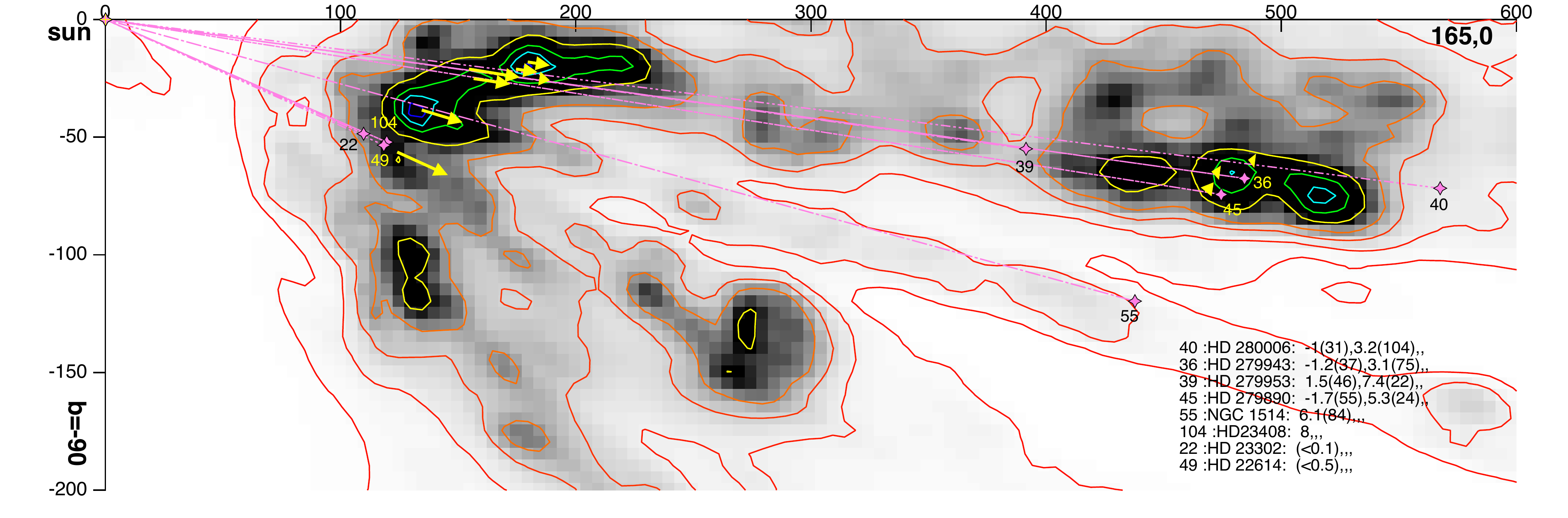}
  \includegraphics[width=0.9\hsize]{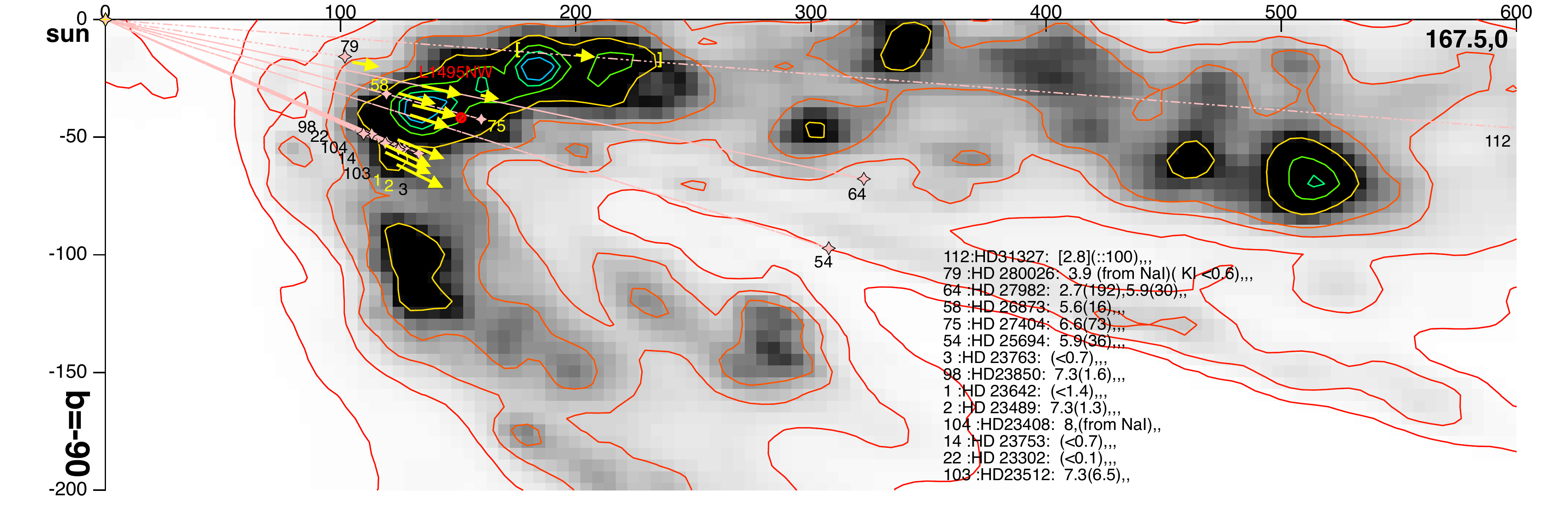}
  \includegraphics[width=0.9\hsize]{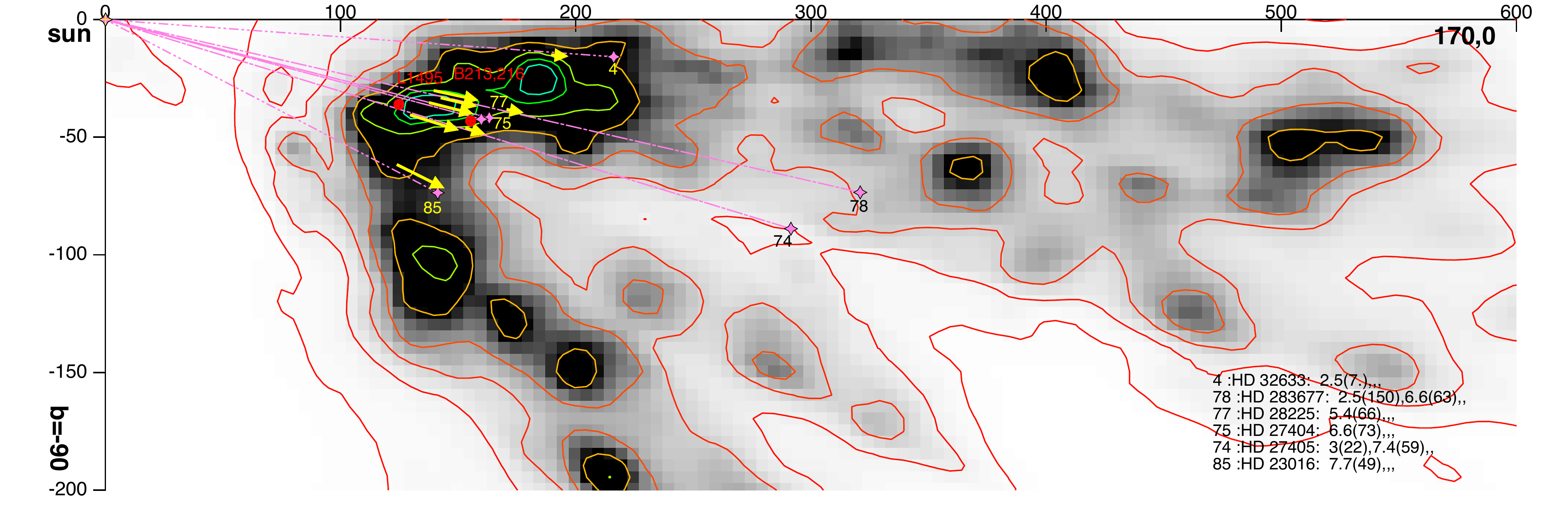}
  \includegraphics[width=0.9\hsize]{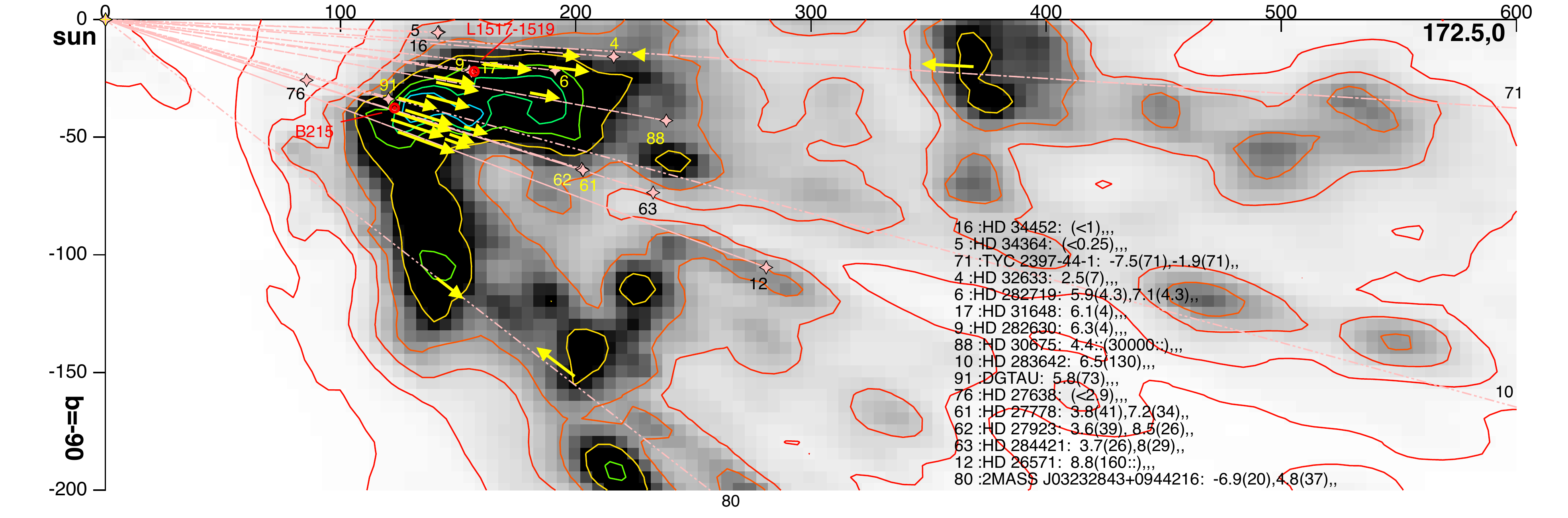}
\caption{Same as Fig \ref{fig:l150}, for longitudes l=165, 167.5, 170 and 172.5\fdeg}
\label{fig:lon2}
\end{figure*}

\begin{figure*}
\centering
  \includegraphics[width=0.9\hsize]{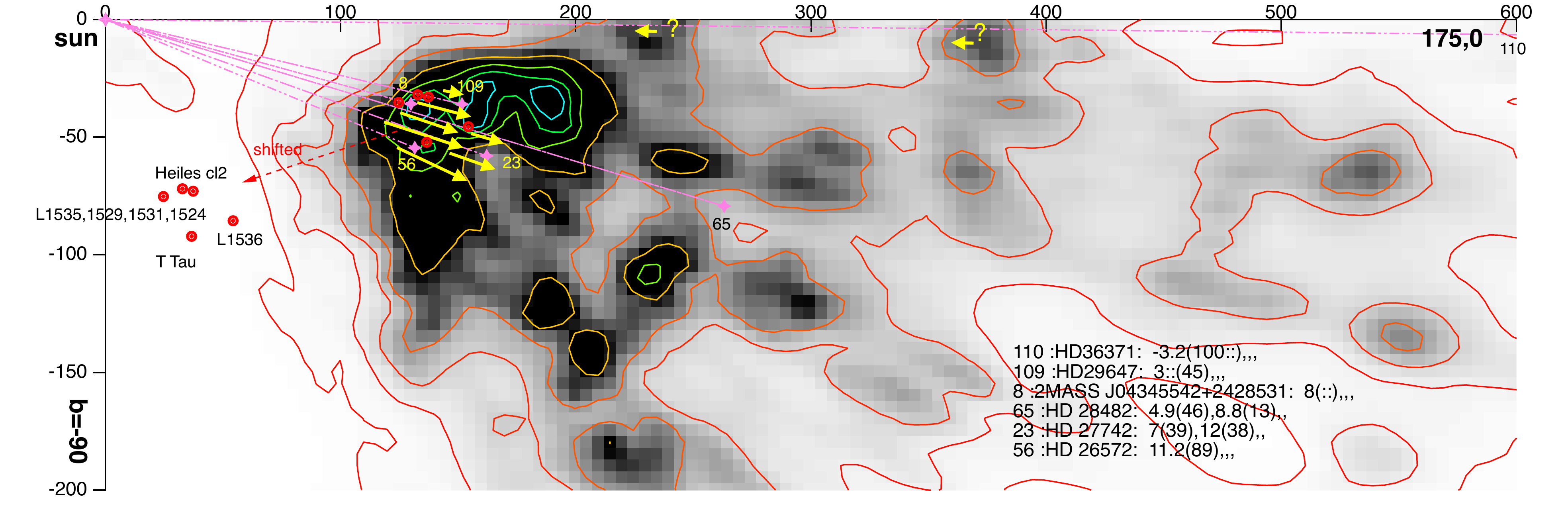}
  \includegraphics[width=0.9\hsize]{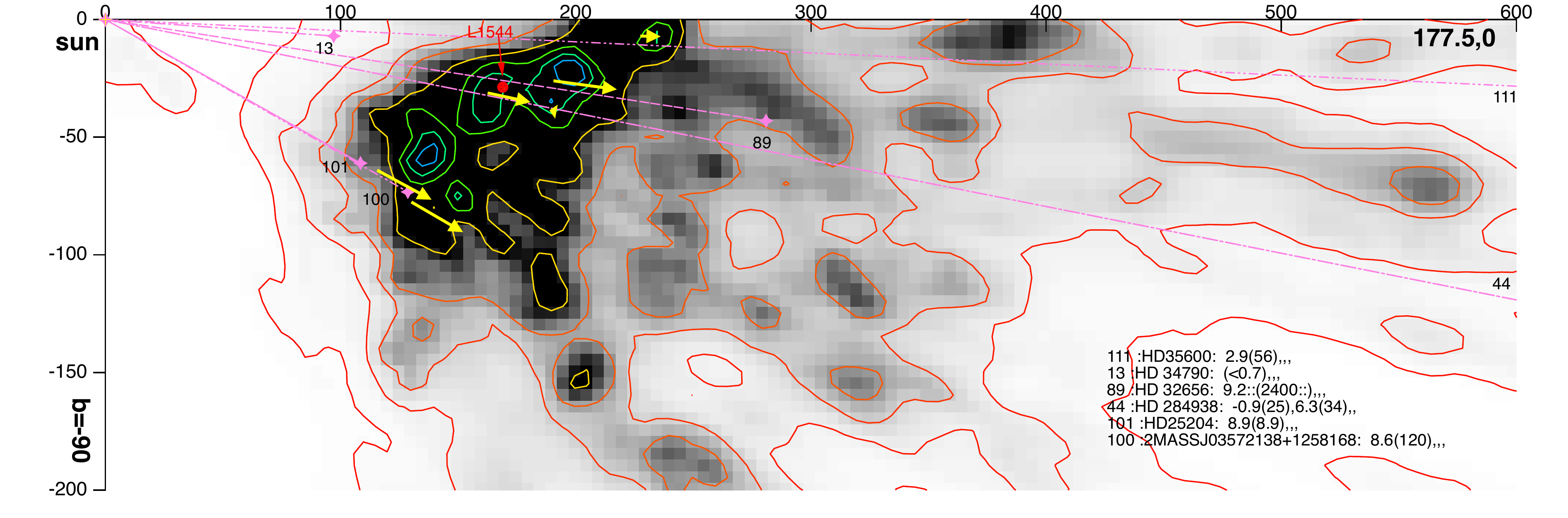}
  \includegraphics[width=0.9\hsize]{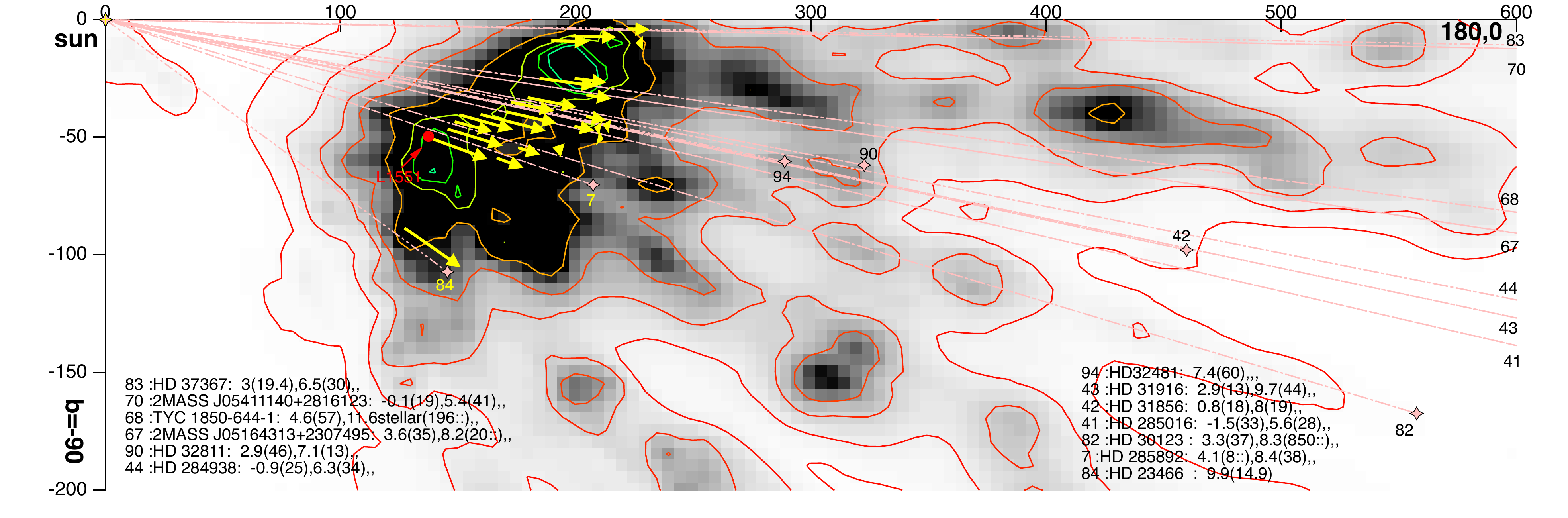}
  \includegraphics[width=0.9\hsize]{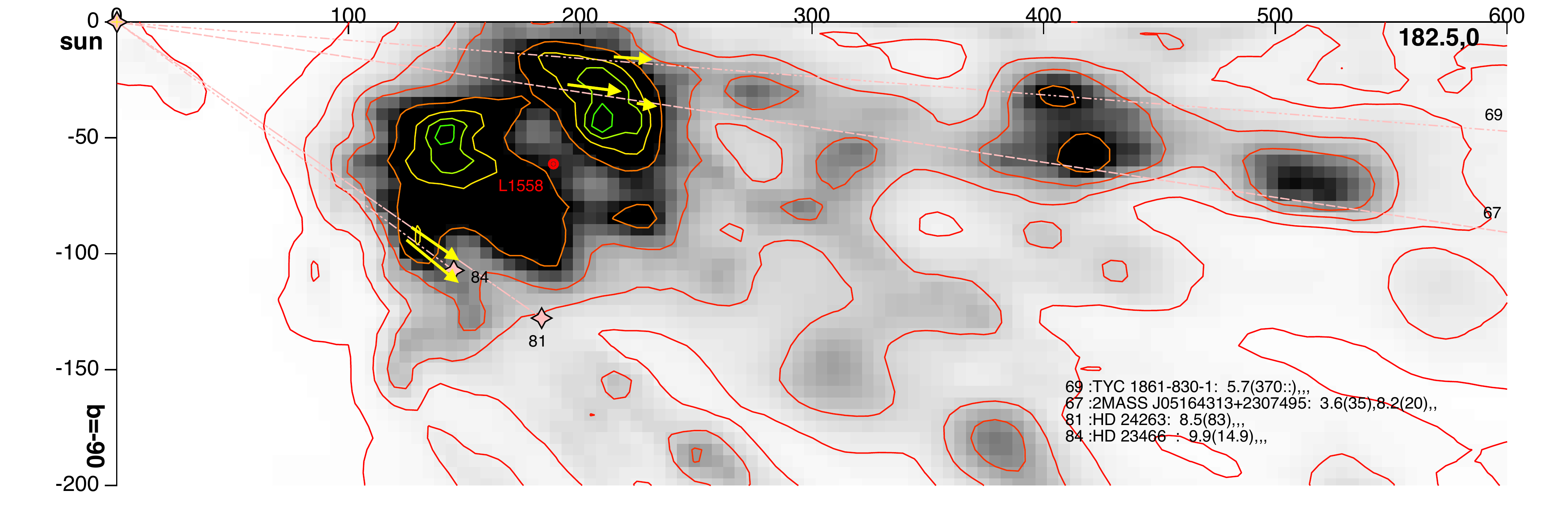}
\caption{Same as Fig \ref{fig:l150}, for longitudes l=175, 177.5, 180 and 182.5\fdeg}
\label{fig:lon3}
\end{figure*}

\section{Conclusions and perspectives}

We have investigated the link between interstellar K\,{\sc i} absorption and dust opacity, and tested the use of K\,{\sc i} absorption data in the context of 3D kinetic tomography of the ISM. To do so, we have obtained and analyzed high-resolution stellar spectra of nearby and mainly early-type stars in the Taurus-Perseus-California area, recorded with the Narval spectrograph at TBL/Pic du Midi. We have developed a new technique based on synthetic atmospheric transmission profiles that allows us to extract a maximum of information from the interstellar K\,{\sc i} absorption doublet and we have applied the new technique to the Narval data as well as to archival data, mainly high-resolution spectra from the Polarbase archive. Atmospheric profiles were downloaded from the TAPAS facility for each observing site and adjusted to the data. During the adjustment the instrumental width and its variation with wavelength was derived. A new interstellar-telluric profile-fitting using Voigt profiles in a classical way AND the previously derived telluric profile was performed, and the radial velocities of the main absorbing interstellar clouds were derived. The adjustment used the measured LSFs, one for each transition of the doublet. We present the results of this profile-fitting for 108 targets, complemented by results from the literature for 8 additional stars. 

 In parallel, with computed an updated 3D distribution of interstellar dust, based on the inversion of a large catalog of extinction measurements for stars distributed at all distances.  The maps were obtained in a similar way to the one based on GAIA/2MASS presented in \cite{Lallement19}, but the inverted dataset was the extinction catalog computed by \cite{Sanders18} as an auxiliary product of their study on stellar ages, augmented by the small compilation of nearby star extinctions used in \cite{Lallement14}. Importantly, instead of an initial, homogeneous, plane-parallel prior, the new prior was the Gaia-2MASS map itself. As a result, in regions that are not covered by the spectroscopic surveys analyzed by \cite{Sanders18}, the recovered 3D distribution is the unchanged prior, i.e. the Gaia/2MASS map, and in areas covered by the surveys there is additional information from the combination of spectral and photometric information. The spatial resolution of the dust maps is limited by the target star density and the correlation length imposed in the inversion. Here, the last iteration was done for a 10 pc wide kernel, and this implies that the recovered dust masses are spread out over volumes on the order of 10 pc or more.  

As a first test of the link between K\,{\sc i} and dust opacity, we compared the equivalent width of the 7699\AA~ absorption with the integral of the differential extinction between the Sun and the target star throughout the new 3D map. The relationship between the two quantities is compared with the results of WH01 for K\,{\sc i} and the H column (HI +2 H$_{2}$) and we obtain  similar correlation coefficients.  This confirms that neutral potassium is tracing both atomic and molecular phases in dense structures. Based on the comparison between the dust cloud distribution and the absorption velocities, we have made a preliminary assignment of radial velocities to the dense structures showing up in the maps. According to the above description, our study had several limitations: the limited number of lines of sight ($\simeq$ 120 targets), the limited spatial resolution of the maps ($\simeq$ 10 pc), the partially arbitrary decomposition of K\,{\sc i} absorption profiles into discrete clouds, leading to minimum velocity difference on the order of 1 to 2 km s$^{-1}$. However, despite of these limitations, we found that, except for very few measurements, it was possible and relatively easy to find a coherent solution for the connections between clouds and velocities, with some continuity in direction and distance for the velocity field. Large-scale spatial gradients were retrieved, and some internal gradients could be determined. We present images of the dust distribution in a series of vertical planes containing the Sun and oriented along Galactic longitudes from 150 to 182.5\fdeg. Proposed velocity assignments and paths to the used target stars are indicated in each image, allowing to visualize the velocity pattern and check the sources of the constraints. We show several comparisons between our maps and velocity assignments with   the locations of molecular cloud and young star clusters and their radial velocities from recent works by \cite{Zucker18, Galli19, Zucker20}. There is a good agreement between for both positions and radial velocities. 

This study shows that it is possible to obtain a first, spatially coherent 3D kinetic tomography of the ISM in out-of-Plane nearby regions, using solely dust maps and K\,{\sc i} absorption data. Such a preliminary tomography can be used as a prior solution for more ambitious, automated kinetic tomography techniques, using dust maps in combination with emission data, and/or with massive amounts of absorption data from future surveys. It shows that K\,{\sc i}  absorptions can also be a very efficient tool to check or validate tomographic results based on other absorption or emission spectral data.  A main advantage of using K\,{\sc i} as a tracer of Doppler velocities is the sharpness of the absorption lines and the resulting optimal disentangling of multiple clouds, at variance with DIBs that are broad. DIBs have already proven to be useful tools for kinetic  measurements in the visible \citep[e.g., ][]{Puspitarini15}, but they are most useful in the infra-red wavelength range due to a better access to target beyond opaque regions \citep{ZasowskiMenard14,Zasowski15,Tchernyshyov18}. In this respect they are superior to K\,{\sc i}. On the other hand, another advantage of K\,{\sc i}, as quoted above, is its association with both atomic and molecular phases, as shown by WH01, and extrapolated here to the dust opacity. For this reason, K\,{\sc i} will be particularly well adapted to kinetic tomography using dust maps and devoted to extended regions consisting of both phases, including dense gas reservoirs devoid of CO.  Ultimately, 3D kinetic tomography may be best achieved through the combination of various tracers, ideally in absorption and emission, and, obviously, using all results directly or indirectly related to Gaia catalogs. In all cases, assigning velocities to structures seen in 3D would allow to link the structures to emission lines that possess particularly rich information on physical and chemical mechanisms at work and to shed light on evolutionary processes.

\longtab{
\begin{longtable}{cccccc}
\caption{\label{tab:targettable} List of targets. Distances are calculated based on parallaxes from Gaia EDR3. (*) means archive data for which a profile-fitting was performed. The second part of the table below the 2 horizontal lines contains data which is published or could be find in other archives, for which no profile-fitting was done. (Na) means that only sodium could be extracted. (7699) means that only the 7699$\AA$  K\,{\sc i}  line could be extracted, otherwise data corresponds to the K\,{\sc i} doublet.}\\
\hline\hline
Star ID & Galactic Longitude & Galactic Latitude &  Distance & Distance Error & Origin \\
\hline 
\endfirsthead
\caption{continued.}\\
\hline\hline
Star ID & Galactic Longitude & Galactic Latitude &  Distance & Distance Error & Origin \\
\hline 
\endhead
\hline
\endfoot
HD 22544&153.75	&-11.34	&356&16	&ESPaDOnS*\\
HD 23642&166.53	&-23.32	&138.3	&0.9&ESPaDOnS*\\
HD 23489&166.37&-23.50&136.1&0.5&ESPaDOnS*\\
HD 23763&166.68	&-23.12	&144.9&4.7&ESPaDOnS*\\
HD 32633&171.16	&-4.19&216.7&1.8&ESPaDOnS*\\
HD 34364&172.77	&-2.23&141.4&0.9&ESPaDOnS*\\
HD 282719&172.79&-6.45&	192.4&0.7&ESPaDOnS*\\
HD 285892&179.21&-18.72	&218.9&0.9&ESPaDOnS*\\
2M J04345542+2428531	&174.32	&-15.40	&134.7&	1.6	&ESPaDOnS*\\
HD 282630&172.68&-8.16	&157.2&0.5&Narval*\\
HD 283642&171.55&-15.35	&653.3&	7.1&ESPaDOnS*	\\
HD 24534&163.08	&-17.14&614&14&ESPaDOnS*	\\
HD 26571&172.42&-20.55&300&10&ESPaDOnS*	\\
HD 34790&176.57	&-4.12&97.3&0.7	&ESPaDOnS*	\\
HD 23753&167.33&-23.83&130.0&1.7&ESPaDOnS*	\\
DG Tau &171.84&-15.76&125.3&2.0&ESPaDOnS*	\\
HD 275877&156.80&-11.90&424.9&5.5&Narval*	\\
HD 34452&172.86	&-2.13&	141.8&2.1&Narval*	\\
HD 31648&173.47	&-7.90&	156.2&1.3&Narval*	\\
HD 24760&157.35	&-10.09&188&15&Narval*	\\
HD 24640&160.47	&-13.97	&423&50&Narval*	\\
HD 23180&160.36	&-17.74	&331&43&Narval*	\\
HD 25940&153.65	&-3.05	&157&7&Narval*	\\
HD 23302&166.18	&-23.85	&120&6&Narval*	\\
HD 27742&175.32	&-19.67	&172.1&1.4&Narval*	\\
HD 24912&160.37	&-13.11	&409&44&Narval*	\\
HD 21540&149.00	&-7.60&375&13&Narval	\\
TYC 2342-639-1&159.44&-21.25&562&7&Narval	\\
BD+45 783&149.97&-8.13&632&22&Narval	\\
HD 22636&149.29&-5.12&586&6&Narval	\\
HD 281160&160.45&-17.89&315.0&2.6&Narval	\\
HD 275941&157.59&-10.67&434&6&Narval	\\
TYC 2864-514-1&158.77&-11.62&664&7&Narval	\\
LS V +43 4&157.03&-5.38&2420&117&Narval	\\
HD 26568&157.02&-5.25&840&17&Narval	\\
HD 276436&161.05&-5.26&866&15&Narval	\\
HD 276453&161.55&-5.41&459.2&4.3&Narval\\
HD 279943&164.31&-7.93&489&6&Narval\\
HD 276449&161.47&-5.17&1620&63&Narval	\\
HD 276454&161.73&-5.37&964&22&Narval	\\
HD 279953&164.77&-7.99&	395.2&3.7&Narval	\\
HD 280006&165.11&-7.20&	572&71&Narval	\\
HD 285016&179.63&-13.00&755&15&Narval	\\
HD 31856&179.53	&-12.02&470.0&5.5&Narval	\\
HD 31916&179.57	&-11.93&704&15&Narval	\\
HD 284938&178.84&-11.23&730&41&Narval	\\
HD 279890&164.54&-8.90&	480.2&4.5&Narval	\\
HD 20326&157.68	&-24.11&244.9&2.1&Narval	\\
HD 20456&157.98	&-23.97&183.1&2.5&Narval	\\
BD+30 540&157.97&-21.24&294.3&2.0&Narval	\\
HD 22614&164.67	&-24.27&129.8&0.6&Narval	\\
HD 278942&159.85&-18.61&383&13&Narval	\\
HD 22780&156.37	&-14.04&194.5&5.1&Narval	\\
BD+30 564&160.77&-18.88&233.0&1.3&Narval	\\
HD 24982&158.42	&-10.75	&255.8&	2.4	&Narval	\\
HD 25694&166.68	&-17.54&322.5&2.2&Narval	\\
NGC 1514&165.53	&-15.29	&453.8&3.5	&Narval	\\
HD 26572&174.69	&-22.50	&142.3&1.0	&Narval	\\
HD 26746&158.58	&-6.45&111.8&0.4&Narval	\\
HD 26873&167.14	&-14.78&123.5&1.3&Narval	\\
HD 27086&162.00	&-9.14&212.9&1.2&Narval	\\
HD 27448&159.72&-5.98&588&10&Narval	\\
HD 27778&172.76&-17.39&212.2&1.3&Narval+UVES	\\
HD 27923&173.38&-17.51&212.9&2.7&Narval	\\
HD 284421&173.44&-17.52	&244.2&1.0&Narval	\\
HD 27982&167.26&-11.86&329.5&7.3&Narval	\\
HD 28482&174.33&-16.76&274.7&1.9&Narval	\\
HD 37752&184.35&-3.65&214.8&1.9&Narval	\\
2M J05164313+2307495&181.34&-8.60&4658.3\footnote{Distance from \cite{BailerJones20}}&1336&Narval\\
TYC 1850-644-1&	179.62&-7.77&673&8&Narval	\\
TYC 1861-830-1&	183.36&-4.49&1158&24&Narval	\\
2M J05411140+2816123&180.06&-1.18&1041&38&Narval\\
TYC 2397-44-1&171.63&-3.58&2636&125&Narval\\
2M J04282605+4129297&160.54&-5.01&1453&46&Narval	\\
HD 281157&160.49&-17.44	&324.9&2.0&Narval	\\
HD 27405&171.02	&-16.95	&304.7&2.5&	Narval	\\
HD 27404&168.71	&-14.85	&165.3&0.7&	Narval	\\
HD 27638&171.51	&-16.74	&89.3&0.7&	Narval	\\
HD 28225&170.74	&-14.35	&168.4&0.7&	Narval	\\
HD 283677&170.37&-12.89	&329.2&2.2&	Narval	\\
HD 280026&167.02&-8.78        &103&23&   Narval	\\
2M J03232843+0944216	&173.19	&-37.88	&1081&21&Narval	\\
HD 32481&180.90&-11.79&465.4&7.5&Narval (7699) \\
HD 281305&162.94&-17.32&273&3&Narval (7699) \\
HD 24263&182.07&-34.87&223.4&3.7&UVES*\\
HD 30123 &180.11&-16.71&582.1&7.2&UVES*\\
HD37367&179.04&-1.03&1312&256&UVES*\\
HD 23466&181.28&-36.39&180.6&3.5&UVES*\\
HD 23016&168.99&-27.51&159.3&2.1&UVES*\\
HD 24398&162.29&-16.69&259&28&UVES*\\
HD 22951&158.92&-16.70&370&18&UVES*\\
HD 30675&173.61&-10.20&242.3&4.4&XSHOOTER*\\
HD 32656&177.18&-8.70&284.1&2.5&XSHOOTER*\\
HD 32811&180.50&-10.85&328.5&3.6&FEROS*	\\
2M J03572138+1258168&177.34&-29.66&147.9&3.5&FEROS (Na, No detected K\,{\sc i})	\\
HD 25204&178.37&-29.38&124.4&6.8&FEROS (Na, No detected K\,{\sc i})  	\\
HD 281159&160.49&-17.80&350.6&9.2&ELODIE (Na, No K\,{\sc i}) 	\\
HD 23512&166.85&-23.95&136.3&2.6&ARCES (7699) 	\\
HD 23408&166.17&-23.51&130.4&5.3&ARCES (Na, No detected K\,{\sc i}) \\
BD+19 441&159.60&-33.54&207.4&1.1&ARCES (7699)	\\
HD 19374&162.98&-34.21&313.7&6.2&ARCES (7699)  	\\
HD 21483&158.87&-21.30&463.2&5.7&ARCES (7699) \\
HD 29647&174.05&-13.35&155.8&1.0&ARCES (7699) 	\\
\hline
\hline
HD 23850&167.01&-23.23&123.2&7.3&Welty \& Hobbs 2001\\
HD 36371&175.77&-0.61&1100.4&241.4&Chaffee\&White 1982\\	
HD 35600&176.76&-2.70&906.9&34.8&Chaffee\&White 1982	\\
HD 31327&168.14&-4.40&942.6&28.9&Chaffee\&White 1982	\\
HD 24432&151.12&-3.50&1139.3&27.1&Chaffee\&White 1982	\\
HD 24131&160.23&-15.14&365.2&10.3&Chaffee\&White 1982	\\
HD 21856&156.32&-16.75&415.5&15.0&Chaffee\&White 1982	\\
HD 21803&150.61&-9.18&543.8&11.0&Chaffee\&White 1982\\	
\end{longtable}
}

\longtab{
\begin{landscape}
\begin{longtable}{ccccccccc}
\caption{\label{tab:velocitytable} Velocity table. Velocities of clouds $V_{1}$, $V_{2}$, $V_{3}$, $V_{4}$ are represented in LSR ($km s^{-1}$). Column densities $N_{1}$, $N_{2}$, $N_{3}$, $N_{4}$ have dimensions E10 $cm^{-2}$. Measurement error is shown in brackets. Similarly to table \ref{tab:targettable}, the second part of the table below the 2 horizontal lines corresponds to data which is published or could be find in other archives, for which no profile-fitting was done}\\
\hline\hline
Star ID&$V_{1}$&$N_{1}$&$V_{2}$&$N_{2}$& $V_{3}$ & $N_{3}$& $V_{4}$ & $N_{4}$\\
\hline
\endfirsthead
\caption{continued.}\\
\hline\hline
Star ID&$V_{1}$&$N_{1}$&$V_{2}$&$N_{2}$& $V_{3}$ & $N_{3}$& $V_{4}$ & $N_{4}$\\
\hline
\endhead
\hline
\endfoot
HD 22544&3.09($\pm$0.2)&17.3($\pm$25.12)&-1.10($\pm$0.4)&17.2($\pm$12.96)&&&&	\\	
HD 23642&&($\leq$0.1)&&&&&&\\	 
HD 23489&&($\leq$1.0)&&&&&&\\ 
HD 23763&&($\leq$0.7)&&&&&&\\  
HD 32633&2.5($\pm$0.09)&7($\pm$1)&&&&&& \\
HD 34364&&($\leq$0.25)&&&&&&\\ 
HD 282719&5.9($\pm$2)&4.3($\pm$2)&7.1($\pm$2)&4.3($\pm$2)&&&& \\
HD 285892&4.1($\pm$0.1)&8::&8.4($\pm$0.6)&38($\pm$10)&&&& \\
2MASS J04345542+2428531	&8.0($\pm$1)&::cool star, no derived column&&&&&& \\
HD 282630&6.26($\pm$0.09)&4($\pm$1)& &&&&& \\
HD 283642&6.45($\pm$0.07)&130($\pm$10)& &&&&& \\
HD 24534&6.12($\pm$0.01)&170($\pm$20)&&&&&& \\
HD 26571&8.83($\pm$0.05)&160($\pm$60)&&&&&& \\
HD 34790&&($\leq$0.7)&&&&&& \\ 
HD 23753&&($\leq$0.8)&&&&&& \\ 
DG Tau &5.8($\pm$1)&73($\pm$10)&&&&&&  \\
HD 275877&-4.93($\pm$0.87)&7($\pm$5)&8.92($\pm$0.43)&	16($\pm$5)&2.06($\pm$0.15)&44($\pm$10)&& \\
HD 34452&&($\leq$1)&&&&&&\\ 
HD 31648&6.1($\pm$0.25)&4($\pm$4)&&&&&&\\
HD 24760&3.3($\pm$1)&1.1($\pm$0.3) &&&&&&\\ 
HD 24640&3.51($\pm$0.16)&8($\pm$1)&11.0($\pm$1)&1::&&&&\\
HD 23180&4.40($\pm$1.0)&26($\pm$5)&7.50($\pm$0.5)&63($\pm$2.5)&&&&\\ 
HD 25940&5.02($\pm$0.05)&11($\pm$3)&&&&&&\\
HD 23302&&($\leq$0.1)&&&&&&\\ 
HD 27742&7.0($\pm$0.5)&39($\pm$2)&12($\pm$1)&38($\pm$5)&&&&\\
HD 24912&5.06($\pm$1.67)&8($\pm$10)&1.34($\pm$0.36)&20($\pm$7)&8.34	($\pm$0.36)&3($\pm$4)&&\\
HD 21540&4.55($\pm$0.21)&29($\pm$8)&-1.69($\pm$0.44)&12($\pm$7)&&&&\\
TYC 2342-639-1&4.23($\pm$0.36)&137($\pm$80)&3.30($\pm$0.46)&42($\pm$6)&&&&\\
BD+45 783&0.1($\pm$0.18)&15.1($\pm$1)&4.1($\pm$1)&15.6($\pm$2)&-7.7 ($\pm1.5$)&4($\pm$1) &&\\
HD 22636&-2.8($\pm$1)&9.6::&-5.9($\pm$1)&10.9::&	1.6($\pm$1)&11.3::	&&\\
HD 281160&7.78($\pm$0.07)&51($\pm$9)&3.34($\pm$0.17)&54($\pm$50)&&&&\\
HD 275941&-8.1($\pm$1)&14::&-4.1($\pm$0.1)&93::&1.4($\pm$0.1)&69::&5.1($\pm$0.6)&15::\\
TYC 2864-514-1&-4.21($\pm$0.41)&13($\pm$7)&	3.88($\pm$0.08)&160($\pm$20)&&&&\\
LS V +43 4	&-9.7($\pm$0.1)&44($\pm$6)&-22.3($\pm$0.2)&12($\pm$4)&-1.8($\pm$0.1)&59($\pm$9)&4.7($\pm$0.1)&8($\pm$5)\\
HD 26568&-16.3($\pm$0.1)&17($\pm$6)&-6.2($\pm$0.3)&31($\pm$11) & -2.0 ($\pm$0.1)&44($\pm$31)&3.7 ($\pm$0.1)& 18 ($\pm$6)\\
HD 276436&-2.59($\pm$1.74)&6($\pm$7)&2.7($\pm$0.47)&20($\pm$8)&&&	&	\\
HD 276453&2.11($\pm$0.11)&18($\pm$4)&&&&&&\\
HD 279943&3.05($\pm$0.75)&75($\pm$30)&-1.21($\pm$1.48)&37($\pm$30)&&&&\\
HD 276449&-9.70($\pm$1.42)&9.7($\pm$10)&-3.47($\pm$1.41)&13($\pm$20)&	2.09($\pm$0.84)&28($\pm$10)&&\\
HD 276454&-5.47($\pm$0.13)&34($\pm$8)&2.83($\pm$0.22)&18($\pm$6)&&&&\\
HD 279953&1.49($\pm$0.18)&46($\pm$9)&7.36($\pm$0.36)&22($\pm$6)&&&&	\\
HD 280006&-0.96($\pm$1.49)&31($\pm$20)&3.18($\pm$0.50)&100($\pm$30)&&&&\\
HD 285016&-1.47($\pm$0.12)&33($\pm$7)&5.63($\pm$0.16)&28($\pm$6)&&&&\\
HD 31856&0.77($\pm$0.25)&18($\pm$5)&7.99($\pm$0.24)&19($\pm$5)&&&&\\
HD 31916&2.89($\pm$0.50)&13($\pm$6)&9.71($\pm$0.18)&44($\pm$9)&&&&	\\
HD 284938&-0.90($\pm$0.14)&25($\pm$6)&6.27($\pm$0.13)&34($\pm$6)&&&&\\
HD 279890&-1.74($\pm$0.11)&55($\pm$7)&5.28($\pm$0.20)&24($\pm$6)&&&&\\
HD 20326&4.2($\pm$3)&6::&&&&&&\\
HD 20456&4.3($\pm$3)&4::&&&&&&\\
BD+30 540&5.03($\pm$0.03)&157::($\pm$10)&&&&&&\\ 
HD 22614&&($\leq$0.5)&&&&&&\\ 
HD 278942&7.73($\pm$0.02)&180($\pm$10)	&&&&&&	\\
HD 22780&1.9($\pm$2)&2.6($\pm$1)&&&&&&\\ 
BD+30 564&4.6($\pm$0.3)&76::($\pm$9)&7.2($\pm$3)&19::&&&&	\\
HD 24982&0.50($\pm$0.06)&30($\pm$20)&&&&&&	\\
HD 25694&5.88($\pm$0.16)&36($\pm$10)&&&&&&	\\
NGC 1514&6.14($\pm$0.03)&84($\pm$10)&&&&&&	\\
HD 26572&11.21($\pm$0.04)&89($\pm$10)&&&&&&	\\
HD 26873&5.60($\pm$0.06)&16($\pm$20)&&&&&&	\\
HD 27086&3.82($\pm$0.18)&23($\pm$20)&&&&&&	\\
HD 27448&-8.51($\pm$0.97)&5.5($\pm$7)	&2.59($\pm$2.40)&11($\pm$10)&	-2.68($\pm$0.83)&31($\pm$20)	&&\\
HD 27778&3.8($\pm$0.5)&41($\pm$1)	&7.2($\pm$0.5)&34($\pm$1)&&&&	\\ 
HD 27923&3.6($\pm$0.1)&39($\pm$11)&8.5($\pm$0.1)&26($\pm$9)&&&&		\\
HD 284421&3.74($\pm$0.14)&26($\pm$30)&7.95($\pm$0.19)&29($\pm$20)&&&&		\\
HD 27982&2.74($\pm$0.6)&192($\pm$1180)&5.94($\pm$5.2)&29.7($\pm$182)&&&&			\\
HD 28482&4.9($\pm$0.5)&46($\pm$5)&8.8($\pm2$)&13($\pm$5)&&&&	\\
HD 37752&9.32($\pm$0.12)&14($\pm$8)&&&&&&	\\
2MASS J05164313+2307495	&3.6($\pm$0.50)&35::&8.2($\pm$0.90)&	21::	&&&&		\\
TYC 1850-644-1	&11.62($\pm$0.11&196($\pm$30)&4.61($\pm$0.22)&57($\pm$10)&&&&			\\
TYC 1861-830-1	&5.7($\pm$0.05	)&370($\pm$60)&&&&&&	\\
2MASS J05411140+2816123	&-0.14($\pm$0.31)&19($\pm$5)&5.36($\pm$0.13)&	41($\pm$7)&&&&\\
TYC 2397-44-1&-7.5($\pm$0.30)&70($\pm$20)&-1.9($\pm$0.32)&71($\pm$20)&&&&\\
2MASS J04282605+4129297	&-12.81($\pm$0.38)&27($\pm$8)&-0.67($\pm$0.23)&	32($\pm$50)&-7.39($\pm$0.88)&8.8($\pm$20)&&			\\
HD 281157&2.69($\pm$0.14)&71($\pm$30)&8.14($\pm$0.03)&160($\pm$10)&&&&	\\
HD 27405&7.43($\pm$0.11)&59($\pm$9)&2.96($\pm$0.89)&22($\pm$100)&&&&		\\
HD27404	&6.58($\pm$0.08)&73($\pm$10)&&&&&&\\
HD 27638&&($\leq$2.9)&&&&&&\\ 
HD 28225&5.41($\pm$0.08)&66($\pm$10)&&&&&&	\\
HD 283677&6.58($\pm$0.13)&63($\pm$30)&2.52($\pm$0.29&150($\pm$40)&&&&	\\
HD 280026&3.9 ($\pm$1) from Na &K\,{\sc i}$\leq$0.6 &&&&&&\\ 
2MASS J03232843+0944216	&-6.9($\pm$0.25)&20($\pm$6)&4.8($\pm$0.2)&	37($\pm$8)&	&&&\\
HD 32481&7.4&60&&&&&&\\
HD 281305&6.4&95::&&&&&&\\
HD 24263 &8.5($\pm$0.05) & 83($\pm$2) &&&&&&\\
HD 30123 &3.3($\pm$1)&37($\pm$5)&8.3($\pm$1)&850::&&&&\\
HD 37367&3.03($\pm$0.04)&19.4($\pm$3)&6.54($\pm$0.03)&30($\pm$3)&&&&	\\
HD 23466&9.9($\pm$0.03)&14.9($\pm$1)&&&&&&	\\
HD 23016&7.72($\pm$0.01)&49($\pm$3)&&&&&&	\\
HD 24398&6.23($\pm$0.01)&160($\pm$20)&&&&&&	\\
HD 22951&6($\pm$1)&33::&1.4($\pm$1)&1.6::&&&&		\\
HD 30675&4.4($\pm$0.20)&30000::&&&&&&	\\
HD 32656&9.2($\pm$0.1)&2400::	&&&&&&	\\
HD 32811&7.07($\pm$0.6)&13($\pm$20)&2.87($\pm$0.4)&46($\pm$30)&&&&	\\
2MASS J03572138+1258168&8.6&120 (Na) & & & & & & \\
HD 25204&8.9&8.9 (Na) & & & & & & \\
HD 281159&8.1&660:: (Na) & & & & & & \\
HD 23512&7.3& 6.5& & & & & & \\
HD 23408&8.0&& & & & && \\
BD+19 441&6.0&11& & & && & \\
HD 19374&-2.9&17&8.8&10&& & & \\
HD 21483&3.05&85& & & && &\\
HD 29647&3.0&45& & & && &\\
\hline
\hline
HD 23850 & 7.3 &1.6 & & & & & & \\
HD 36371&-3.2&100::& & & && &\\
HD 35600&2.9&56& & & && &\\
HD 31327&2.83&100 &&& && &\\
HD 24432&-8.40 &162 &2.80 & && &&\\
HD 24131&7.88 &32 & & && &&\\
HD 21856&-0.22 &30 & & && &&\\
HD 21803&2.64&30 & & && &&\\
\end{longtable}
\end{landscape}
}

\begin{acknowledgements}
R.L. deeply thanks the Pic du Midi TBL staff for his efficiency and assistance. A.I. wants to acknowledge the support of Ministry of Science and Higher Education of the Russian Federation under the grant 075-15-2020-780 (N13.1902.21.0039). J.L. Vergely acknowledges support from the EXPLORE project. EXPLORE has received funding from the European Union Horizon 2020 research and innovation programme under grant agreement No 101004214. C.H. acknowledges support from CNES with a post-doctoral research grant. This research has made use of the SIMBAD database, operated at CDS, Strasbourg, France.
\end{acknowledgements}

\bibliographystyle{aa}
\bibliography{Taurus_rev}

\end{document}